\documentclass[10pt]{article}
\usepackage{amsmath,amsfonts,url,epsfig}
\usepackage{graphicx,psfrag,epsf,authblk}
\usepackage{enumerate}
\usepackage{times}
\usepackage{bm}
\usepackage{algorithm}
\usepackage{algpseudocode}
\def\T{{ \mathrm{\scriptscriptstyle T} }}
\def\mR{\mathbb{R}}

\def\tr{\mbox{tr}}

\def\vec{\mbox{vec}}

\def\diag{\mbox{diag}}

\DeclareMathOperator*{\argmin}{arg\,min} 
\newtheorem{theorem}{Theorem}

\newtheorem{proposition}{Proposition}
\newtheorem{remark}{Remark}
\usepackage[round,authoryear]{natbib} 
\begin{document}
\title{An efficient ADMM algorithm for high dimensional precision matrix estimation via penalized quadratic loss}
\author[a]{Cheng Wang \thanks{chengwang@sjtu.edu.cn}}
\author[b]{Binyan Jiang \thanks{by.jiang@polyu.edu.hk}}
\affil[a]{School of Mathematical Sciences, MOE-LSC,Shanghai Jiao Tong University,  Shanghai 200240, China.}
\affil[b]{Department of Applied Mathematics, The Hong Kong Polytechnic University,
	Hung Hom, Kowloon, Hong Kong.}
	\date{\today}
\maketitle
	
	\begin{abstract}
The estimation of high dimensional precision matrices has been a central topic in statistical learning. However, as the number of parameters scales quadratically with the dimension $p$, many state-of-the-art methods do not scale well to solve problems with a very large $p$. In this paper, we propose a very efficient algorithm for precision matrix estimation via penalized quadratic loss functions. Under the high dimension low sample size setting, the  computation complexity of our algorithm is linear in both the sample size and the number of parameters. Such a computation complexity is in some sense optimal, as it is the same as the complexity needed for computing the sample covariance matrix. Numerical studies show that our algorithm is much more efficient than other state-of-the-art methods when the dimension $p$ is very large.

		Key words and phrases: ADMM, High dimension, Penalized quadratic loss, Precision matrix.
	\end{abstract}

\section{Introduction}
\label{sec:intro}

Precision matrices play an important role in statistical learning and data analysis. 
On the one hand, estimation of the precision matrix is oftentimes required in various  statistical analysis. On the other hand,   
under Gaussian assumptions, the precision matrix has been widely used to study the conditional independence among the random variables. Contemporary applications usually require fast methods for estimating a very high dimensional precision matrix \citep{meinshausen2006high}.  
Despite recent advances, estimation of the precision matrix remains challenging when the dimension $p$ is very large, owing to the fact that the number of parameters to be estimated is of order $O(p^2)$. For example, in the Prostate dataset we are studying in this paper, 6033 genetic activity measurements are recorded for 102 subjects. The precision matrix to be estimated is of dimension $6033 \times 6033$, resulting in more than 18 million parameters.

A well-known and popular method in precision matrix estimation is the graphical lasso \citep{yuan2007model, banerjee2008model, friedman2008sparse}. Without loss of generality, assume that $X_1,\cdots,X_n$ are i.i.d. observations from a $p$-dimensional Gaussian distribution with mean ${\bf 0}$ and covariance matrix $\Sigma$. To estimate the precision matrix $\Omega^\ast=\Sigma^{-1}$, the graphical lasso seeks the minimizer of the following  $\ell_1$-regularized negative log-likelihood:

\begin{align} \label{glasso}
\tr(S \Omega)-\log|\Omega|+\lambda \|\Omega\|_1,
\end{align}
over the set of positive definite matrices. Here $S$ is the sample covariance matrix, $\|\Omega\|_1$ is the element-wise $\ell_1$ norm of $\Omega$, and $\lambda \geq 0$ is the tuning parameter. Although \eqref{glasso} is constructed based on Gaussian likelihood, it is known that the graphical lasso also works for non-Gaussian data \citep{ravikumar2011high}. 
Many algorithms have been developed to solve the graphical lasso. \cite{friedman2008sparse} proposed a coordinate descent procedure and \cite{boyd2011distributed} provided an alternating direction method of multipliers (ADMM) algorithm for solving \eqref{glasso}. In order to obtain faster convergence for the iterations, second order methods and proximal gradient algorithms on the dual problem are also well developed; see for example \cite{hsieh2014quic}, \cite{dalal2017sparse}, and the references therein. However, eigen-decomposition or calculation of the determinant of a $p \times p$ matrix is inevitable in these algorithms, owing to the matrix determinant term in \eqref{glasso}. Note that the computation complexity of eigen-decomposition or matrix determinant is of order $O(p^3)$. Thus, the computation time for these algorithms will scale up cubically in $p$. 

Recently,  \cite{zhang2014sparse} and \cite{liu2015fast}  proposed to estimate $\Omega^\ast$ by some trace based quadratic loss functions. Using the Kronecker product and matrix vectorization, our interest is to estimate,
 
\begin{align}\label{form2}
\vec{(\Omega^\ast)}=\vec(\Sigma^{-1})=\vec(\Sigma^{-1}\cdot I_p \cdot  I_p)=(I_p \otimes \Sigma) ^{-1} \vec(I_p),
\end{align}
or equivalently, 

\begin{align}\label{form1}
\vec{(\Omega^\ast)}=\vec(\Sigma^{-1})=\vec(I_p \cdot I_p \cdot \Sigma^{-1})=(\Sigma \otimes I_p) ^{-1} \vec(I_p),
\end{align}
where $I_p$ denotes the identity matrix of size $p$, and $\otimes$ is the Kronecker product. Motivated by \eqref{form2} and the LASSO \citep{tibshirani1996regression}, a natural way to estimate $\beta^\ast=\vec{(\Omega^\ast)}$ is 
\begin{align}\label{liu}
\argmin_{\tiny {\beta \in \mR^{p^2 }}}  \frac{1}{2} \beta^\T (I_p \otimes S) \beta -\beta^\T\vec(I_p)+\lambda \|\beta\|_1. 
\end{align}
To obtain a symmetric estimator we can use both \eqref{form2}  and \eqref{form1}, and estimate $\beta^\ast$ by 
\begin{align}\label{zou}
\argmin_{\tiny {\beta \in \mR^{p^2 }}}  \frac{1}{4} \beta^\T (S \otimes I_p+I_p \otimes S  ) \beta -\beta^\T\vec(I_p)+\lambda \|\beta\|_1. 
\end{align}
 Denoting $\beta=\vec(\Omega)$, \eqref{liu} can be written in matrix notation as
\begin{align} \label{quad}
\hat{\Omega}_1=\argmin_{\tiny {\Omega \in \mR^{p\times p}}} \frac{1}{2}\tr(\Omega^T S \Omega  )-\tr(\Omega)+\lambda \|\Omega\|_1.
\end{align}  
Symmetrization can then be applied to obtain a final estimator.
The loss function
\begin{align*} 
L_1(\Omega):= \frac{1}{2}\tr(\Omega^\T S \Omega  )-\tr(\Omega),
\end{align*}
is used in \cite{liu2015fast} and they proposed a column-wise estimation approach called SCIO. \cite{lin2016estimation}  obtained these quadratic losses from a more general score matching principle. Similarly, the matrix form of \eqref{zou} is
\begin{align} \label{squad}
\hat{\Omega}_2=\argmin_{\tiny {\Omega \in \mR^{p\times p}}}  \frac{1}{4}\tr(\Omega^\T S \Omega )+ \frac{1}{4}\tr(\Omega S \Omega ^\T )-\tr(\Omega)+\lambda \|\Omega\|_1.
\end{align}
The loss function 
\begin{align*}
L_2(\Omega)=\frac{1}{2}\{L_1(\Omega^\T)+L_1(\Omega) \}=\frac{1}{4}\tr(\Omega S \Omega ^\T)+\frac{1}{4}\tr(\Omega^\T S \Omega)-\tr(\Omega),
\end{align*}
is equivalent to the D-trace loss proposed by \cite{zhang2014sparse}, owing to the fact that $L_2(\Omega)$ naturally force the solution to be symmetric.

In the original papers by \cite{zhang2014sparse} and  \cite{liu2015fast}, the authors have established consistency results for the estimators \eqref{quad} and \eqref{squad}  and have shown that their performance is comparable to the graphical lasso. As can be seen in the vectorized formulation \eqref{form2} and \eqref{form1}, the loss functions $L_1(\Omega)$ and $L_2(\Omega)$ are quadratic in $\Omega$. In this note,  we propose efficient ADMM algorithms for the estimation of the precision matrix via these  {qua}dratic loss functions.
In Section 2, we show that under the quadratic loss functions, explicit solutions can be obtained in each step of the ADMM algorithm. In particular, we derive explicit formulations for the inverses of $(S+\rho I)$ and $(2^{-1}S \otimes I+2^{-1}I \otimes S+\rho I)$ for any given $\rho>0$, from which we are able to solve \eqref{quad} and \eqref{squad}, or equivalently  \eqref{liu} and \eqref{zou}, with computation complexity of order $O(np^2)$. Such a rate is in some sense optimal, as the complexity for computing $S$ is also of order $O(np^2)$. Numerical studies are provided in Section 3 to demonstrate the computational efficiency and the  estimation accuracy of our proposed algorithms.  An R package ``EQUAL" has been developed to implement our methods and is available at \url{https://github.com/cescwang85/EQUAL}, together with all the simulation codes. All technical proofs are relegated to the Appendix section.

\section{Main Results}
For any real matrix $M$, we use $\|M\|_2=\sqrt{\tr(MM^T)}$ to denote the Frobenius norm, $\|M\|$ to denote the spectral norm, i.e., the square root of the largest eigenvalue of $M^\T M$, and $\|M\|_\infty$ to denote the matrix infinity norm, i.e., the element of $M$ with largest absolute value.

We consider estimating the precision matrix via
\begin{align}\label{equal_loss}
\argmin_{\tiny {\Omega \in \mR^{p\times p}}}  L(\Omega)+\lambda \|\Omega\|_1,
\end{align}
where $L(\Omega)$ is the quadratic loss function $L_1(\Omega)$ or $L_2(\Omega)$ introduced above. The augmented Lagrangian is
\begin{align*}
L_a(\Omega,A,B)= L(\Omega)+\rho/2 \|\Omega-A+B\|_2^2+\lambda \|A\|_1,
\end{align*}
where $\rho>0$ is the step size in the ADMM algorithm. By \cite{boyd2011distributed}, the alternating iterations are
\begin{align*}
\Omega^{k+1}&=\argmin_{\tiny {\Omega \in \mR^{p\times p}}}  L_a(\Omega,A^k,B^k),\\
A^{k+1}&=\argmin_{\tiny {A \in \mR^{p\times p}}}  L_a(\Omega^{k+1},A,B^k)=\mbox{soft}(\Omega^{k+1}+B^k,\lambda/\rho),\\
B^{k+1}&=\Omega^{k+1}-A^{k+1}+B^k,
\end{align*}
where $\mbox{soft}(A,\lambda)$ is an element-wise soft thresholding operator.  
Clearly the computation complexity will be dominated by the update of $\Omega^{k+1}$, which amounts to solving the following problem:
\begin{align}\label{of}
\argmin_{\tiny {\Omega \in \mR^{p \times p}}}  L(\Omega)+\rho/2 \|\Omega-A^{k}+B^k\|_2^2.
\end{align}
From the convexity of the objective function, the solution of \eqref{of} satisfies   
 \begin{align*}
 L'(\Omega)+\rho (\Omega-A^{k}+B^k)=0. 
 \end{align*}
Consequently, for the estimation \eqref{quad} and \eqref{squad}, we need to solve the following equations respectively,
\begin{align}
&S \Omega+\rho \Omega =I_p+\rho (A^{k}-B^k),\label{eq1}\\
&2^{-1}S \Omega +2^{-1}\Omega S+\rho \Omega =I_p+\rho (A^{k}-B^k). \label{eq2}
\end{align}
By looking at \eqref{eq1} and the vectorized formulation of \eqref{eq2} (i.e. equation \eqref{zou}), we immediately have that, in order to solve \eqref{eq1} and \eqref{eq2}, we need to compute the inverses of $(S +\rho I)$ and $(2^{-1}S \otimes I+2^{-1}I \otimes S+\rho I)$. The following proposition provides explicit expressions for these inverses. 
\begin{proposition} \label{prop0}
	Write the decomposition of $S$ as $S=U \Lambda U^\T$ where $U \in \mR ^{p \times m},~m=\min(n,p),~U^ \T U=I_m$ and $\Lambda=\diag\{\tau_1,\cdots,\tau_m\},~\tau_1,\dots,\tau_m\geq0$. For any $\rho>0$, we have
	\begin{align}\label{f1}
	(S+\rho I_p)^{-1}=\rho^{-1} I_p-\rho^{-1} U \Lambda_1 U ^\T,
	\end{align}
	and 
	\begin{align}\label{f2}
	&(2^{-1}S \otimes I_p+2^{-1}I_p \otimes S+\rho I_{p^2})^{-1}\nonumber\\
	=&\rho^{-1} I_{p^2}-\rho^{-1} (U \Lambda_2 U^\T) \otimes I_p-\rho^{-1} I_p \otimes (U \Lambda_2 U^\T)\nonumber\\
	&+\rho^{-1} (U \otimes U) \diag\{\vec(\Lambda_3)\}(U^\T \otimes U^\T ), 
	\end{align}
where 
	\begin{align*}
	\Lambda_1=&\diag\left\{\frac{\tau_1}{\tau_1+\rho},\cdots,\frac{\tau_m}{\tau_m+\rho}\right\}, ~~\Lambda_2=\diag\left\{\frac{\tau_1}{\tau_1+2 \rho},\cdots,\frac{\tau_m}{\tau_m+ 2\rho}\right\},\\ \Lambda_{3}=&\left\{\frac{\tau_i \tau_j(\tau_i+\tau_j+4\rho)}{(\tau_i+2\rho)(\tau_j+2\rho)(\tau_i+\tau_j+2\rho)}\right\}_{m \times m}.
	\end{align*}
\end{proposition}
Using the explicit formulas of Proposition \ref{prop0}, we can solve 
\eqref{eq1} and \eqref{eq2} efficiently.
\begin{theorem} \label{prop1}
	For a given $\rho>0$,
	\begin{itemize} 
		\item[(i)] the solution to the equation 
		$
		S \Omega +\rho \Omega=C 
		$
		is unique and is given as 
		$
		\Omega=\rho^{-1}C-\rho^{-1}U \Lambda_1 U  ^\T C$; 
		\item[(ii)] the solution to the equation 
		$
		2^{-1}S \Omega +2^{-1}\Omega S+\rho \Omega=C 
		$
		is unique and is given as 
$
		\Omega=\rho^{-1}C-\rho^{-1}C U \Lambda_2 U  ^\T- \rho^{-1}U \Lambda_2 U ^\T C+\rho^{-1} U \{\Lambda_3 \circ (U ^\T C U)\} U ^\T,
$
where $\circ$ denotes the Hadamard product.
	\end{itemize}	
\end{theorem}
Note that when $S$ is the the sample covariance matrix, 
$U$ and $\Lambda$ can be obtained from the thin singular value decomposition (Thin SVD) of $X=(X_1,\ldots, X_n)$ whose complexity is of order $O(mn p)$. On the other hand, the solutions obtained in Theorem \ref{prop1} involve only elementary matrix operations of $p\times m$ and $m\times m$ matrices and thus the complexity for solving \eqref{eq1} and \eqref{eq2} can be seen to be of order $O(mnp+mp^2)=O(n p^2)$. 

Based on Theorem \ref{prop1} we next provide an efficient ADMM algorithm for solving \eqref{equal_loss}.
For notation convenience, we shall use the term ``EQUAL" to denote our proposed \underline{E}fficient ADMM algorithm via the \underline{QUA}dratic \underline{L}oss $L(\Omega)=L_1(\Omega)$, and similarly, use ``EQUALs" to denote the estimation based on the symmetric quadratic loss $L(\Omega)=L_2(\Omega)$. The algorithm is given as follows.
 \begin{algorithm}[H]\small 
	\caption{{E}fficient ADMM algorithm via the  {qua}dratic {l}oss $L_1(\Omega)$ or $L_2(\Omega)$.}
	\begin{algorithmic}[1]
			\item[Initialization:]
		\State  Thin SVD of $X$ to obtain $S=U\Lambda U^\T$ where $U \in \mR ^{p \times m},~m=\min(n,p),~U^ \T U=I_m$ and $\Lambda=\diag\{\tau_1,\cdots,\tau_m\},~\tau_1,\dots,\tau_m\geq0$. 
	\State Define 	\begin{align*}
	\Lambda_1=&\diag\left\{\frac{\tau_1}{\tau_1+\rho},\cdots,\frac{\tau_m}{\tau_m+\rho}\right\},~~ \Lambda_2=\diag\left\{\frac{\tau_1}{\tau_1+2 \rho},\cdots,\frac{\tau_m}{\tau_m+ 2\rho}\right\},\\ \Lambda_{3}=&\left\{\frac{\tau_i \tau_j(\tau_i+\tau_j+4\rho)}{(\tau_i+2\rho)(\tau_j+2\rho)(\tau_i+\tau_j+2\rho)}\right\}_{m \times m}.
	\end{align*}
	\State  Start from $k=0$, $B^0={I}_p$ and $A^{0}=I_p$.
		\item[Iteration:] 	
			\State  $k=k+1$, $C= I_p+\rho (A^{k-1}-B^{k-1})$.
			\State 	Update $\Omega^k=\rho^{-1}C-\rho^{-1}U \Lambda_1 U  ^\T C$ when Method=``EQUAL", or Update $\Omega^k=\rho^{-1}C-\rho^{-1}C U \Lambda_2 U  ^\T- \rho^{-1}U \Lambda_2 U ^\T C+\rho^{-1} U \{\Lambda_3 \circ (U ^\T C U)\} U ^\T$ when Method=``EQUALs";
		\State  Update $A^k=\mbox{soft}(\Omega^{k}+B^{k-1},\lambda/\rho)$.
		\State Update $B^k=\Omega^{k}-A^{k}+B^{k-1}$.
		\State 	Repeat steps 4-7 until convergence. 
	
		\item[Output:] Return $\hat{\Omega}=(\omega_{ij})_{p\times p}$ where $\omega_{ij}=A_{ij}^kI\{|A_{ij}^k|<|A^k_{ji}|\}+A_{ji}^kI\{|A_{ij}^k|\geq|A^k_{ji}|\}$ when Method=``EQUAL", 
		or return $\hat{\Omega}=A^k$  when Method=``EQUALs".
	\end{algorithmic}
\end{algorithm}

The following remarks provide further discussions on our approach. 
\begin{remark}
	Generally, we can specify different weights for each element and consider the estimation 
	\begin{align*}
	\hat{\Omega}_k=\argmin_{\tiny {\Omega \in \mR^{p\times p}}}  L_k(\Omega)+\lambda \|W \circ \Omega\|_1,k=1,2,
	\end{align*}
	where $W=(w_{ij})_{p \times p},~w_{i,j} \geq 0$.
	For example, 
	\begin{itemize}
		\item Setting $w_{ii}=0$ and $w_{i,j}=1,~i \neq j$ where the diagonal elements are left out of penalization;
		\item Using the local linear approximation \citep{zou2008one}, we can set
		$
		W=\{ p'_\lambda(\hat{\Omega}_{ij})\}_{p\times p},
		$
		where  $\hat{\Omega}=(\hat{\Omega}_{ij})_{p\times p}$ is a LASSO solution and $p_\lambda(\cdot)$ is a general penalized function such as SCAD or MCP. 
	\end{itemize}
	The ADMM algorithm will be the same as the $\ell_1$ penalized case, except that the $A^{k+1}$ related update is replaced by a element-wise soft thresholding with different thresholding parameters. More details will be provided in Section 3.3 for better elaboration. 
\end{remark}
\begin{remark}
	Compared with the ADMM algorithm given in \cite{zhang2014sparse}, our update of $\Omega^{k+1}$ only involves matrix operations of some $p\times m$ and $m\times m$ matrices, while matrix operations on some $p\times p$ matrices are required in \cite{zhang2014sparse}; see for example Theorem 1 in \cite{zhang2014sparse}. Consequently, we are able to obtain the order $O(np^2)$ in these updates while \cite{zhang2014sparse} requires $O((n+p)p^2)$. Our algorithm thus scales much better when $n \ll p$.
\end{remark}

\begin{remark}
For the graphical lasso, we can also use ADMM \citep{boyd2011distributed} to implement the minimization where the loss function is 
	$
	L(\Omega)=\tr(S \Omega)-\log{|\Omega|}.
	$
	The update for $\Omega^{k+1}$ is obtained by solving 
	$\rho \Omega-\Omega^{-1}=\rho (A^{k}-B^k)-S.$
	Denote the eigenvalue decomposition of $\rho (A^{k}-B^k)-S$ as $Q^\T \Lambda_0 Q$ where  $\Lambda_0=\diag\{a_1,\cdots,a_p\}$, we can obtain a closed form solution,
	\begin{align*}
	\Omega^{k+1}=Q^\T \diag\left\{\frac{a_1+\sqrt{a_1^2+4\rho}}{2 \rho},\cdots,\frac{a_p+\sqrt{a_p^2+4\rho}}{2 \rho}\right\} Q.
	\end{align*}
	Compared with the algorithm based on quadratic loss functions, the computational complexity is dominated by the eigenvalue decomposition of $p \times p$ matrices which is of order $O(p^3)$.
\end{remark}

\begin{remark}
%
A potential disadvantage of our algorithm is the loss of positive definiteness. Such an issue was also encountered in other approaches, such as the SCIO algorithm in \cite{liu2015fast}, the CLIME  algorithm in \cite{cai2011constrained}, and thresholding based estimators \citep{bickel2008covariance}.  From the perspective of optimization, it is ideal to find the solution over the convex cone of positive definite matrices. However, this could be costly, as we would need to
guarantee the solution in each iteration to be positive definite. 
By relaxing the positive definite constraint, much more efficient algorithms can be developed. In particularly, our algorithms turn out to be computationally optimal as the complexity is the same as that for computing a sample covariance matrix. On the other hand, the positive definiteness of the quadratic loss based estimators can still be obtained with statistical guarantees  or by further refinements. More specifically, from Theorem 1 of \cite{liu2015fast} and Theorem 2 of  \cite{zhang2014sparse}, the   estimators are consistent under mild sparse assumptions, and will be positive definite with probability tending to 1. 
 In the case when an estimator is not positive definite, a refinement procedure  which pulls the negative eigenvalues of the estimator to be positive can be conducted to fulfill the positive definite requirement. As shown in \cite{cai2012optimal}, the refined estimator will still be consistent in estimating the precision matrix. 

\end{remark}

\section{Simulations}
In this section, we conduct several simulations to illustrate the efficiency and estimation accuracy of our proposed methods. We consider the following three precision matrices:
\begin{itemize}
	\item Case 1: asymptotic sparse matrix:
	\begin{align*}
	\Omega_1=(0.5^{|i-j|})_{p \times p};
	\end{align*}
	\item Case 2: sparse matrix:
		\begin{align*}
\Omega_2=\Omega_1^{-1}=\frac{1}{3}\begin{pmatrix}
	4&-2&&& \\
	-2&5&-2&&\\
	&\ddots&\ddots&\ddots& & \\
	&&-2&5&-2\\
	&&&-2&4\\
	\end{pmatrix};
	\end{align*}
	\item Case 3: block matrix with different weights: 
	\begin{align*}
	\Omega_3=\diag\{w_1 \Omega_0, \cdots, w_{p/5}  \Omega_0\},
	\end{align*}
	where $\Omega_0 \in \mR^{5 \times 5}$ has off-diagonal entries equal to 0.5 and diagonal 1. The weights $w_1,\cdots, w_{p/5}$ are generated from the uniform distribution on $[0.5,5]$, and rescaled to have mean 1.
\end{itemize}
 For all of our simulations, we set the sample size $n=200$ and generate the data $X_1,\cdots,X_n$ from $N(0,\Sigma)$ with $\Sigma=\Omega_i^{-1},~i=1,2,3$.  

\subsection{Computation time}
For comparison, we consider the following competitors:
\begin{itemize}
	\item CLIME \citep{cai2011constrained} which is implemented by the R package ``fastclime" \citep{pang2014fastclime};
	\item glasso \citep{friedman2008sparse} which is implemented by the R package ``glasso";
	\item BigQuic \citep{hsieh2013big} which is implemented by the R package ``BigQuic";
	\item glasso-ADMM which solves the glasso by ADMM \citep{boyd2011distributed};
	\item SCIO \citep{liu2015fast} which is implemented by the R package ``scio";
	\item D-trace \citep{zhang2014sparse} which is implemented using the ADMM algorithm provided in the paper.
\end{itemize}
Table \ref{tab1} summaries the computation time in seconds based on 100 replications where all methods are implemented in R with a PC with 3.3 GHz Intel Core i7 CPU and 16GB memory. For all the methods, we solve a solution path corresponding to 50 $\lambda$ values ranging from $\lambda_{max}$ to $\lambda_{max} \sqrt{\log{p}/n} $. Here $\lambda_{max}$ is the maximum absolute elements of the sample covariance matrix. Although the stopping criteria is different for each method, we can see from Table \ref{tab1} the computation advantage of our methods. 
In particularly, our proposed algorithms are much faster than the original quadratic loss based methods ``SCIO" or ``D-trace" for large $p$. In addition, we can roughly observe that the required time increases quadratically in $p$ in our proposed algorithms.   

\begin{table}[!htbp] 
	\caption{The average computation time (standard deviation) of solving a solution path for the precision matrix estimation. } 
	\label{tab1} 
		\resizebox{\textwidth}{!}{%
	\begin{tabular}{@{\extracolsep{5pt}} cccccc} 
		\\[-1.8ex]\hline 
		\hline \\[-1.8ex] 
		& p=100 & p=200 & p=400 & p=800 & p=1600 \\ 
		\hline \\[-1.8ex] 
		&\multicolumn{5}{c}{Case 1:  $\Omega=\Omega_1$}\\  \\[-1.8ex] 
	CLIME & 0.390(0.025) & 2.676(0.101) & 15.260(0.452) & 117.583(4.099) & 818.045(11.009) \\ 
glasso & 0.054(0.009) & 0.295(0.052) &  1.484(0.233) &   8.276(1.752) &  45.781(12.819) \\ 
BigQuic & 1.835(0.046) & 4.283(0.082) & 11.630(0.368) &  37.041(1.109) & 138.390(1.237) \\ 
glasso-ADMM & 0.889(0.011) & 1.832(0.048) &  5.806(0.194) &  21.775(0.898) &  98.317(2.646) \\ 
SCIO & 0.034(0.001) & 0.238(0.008) &  1.696(0.041) &  12.993(0.510) & 106.588(0.271) \\ 
EQUAL & 0.035(0.001) & 0.184(0.008) &  0.684(0.045) &   3.168(0.241) &  15.542(0.205) \\ 
D-trace & 0.034(0.002) & 0.215(0.010) &  1.496(0.107) &  11.809(1.430) & 118.959(1.408) \\ 
EQUALs & 0.050(0.002) & 0.294(0.014) &  0.903(0.053) &   3.725(0.257) &  18.860(0.231) \\ 
		\hline \\[-1.8ex] 
		&\multicolumn{5}{c}{Case 2: $\Omega=\Omega_2$} \\  \\[-1.8ex] 
CLIME & 0.361(0.037) & 2.583(0.182) & 14.903(0.914) & 114.694(2.460) & 812.113(16.032) \\ 
glasso & 0.095(0.012) & 0.576(0.069) &  2.976(0.397) &  15.707(2.144) &  93.909(16.026) \\ 
BigQuic & 2.147(0.040) & 5.360(0.099) & 15.458(0.347) &  51.798(1.059) & 186.025(3.443) \\ 
glasso-ADMM & 0.949(0.016) & 1.976(0.056) &  5.710(0.161) &  19.649(0.428) & 123.950(6.130) \\ 
SCIO & 0.039(0.001) & 0.263(0.007) &  1.762(0.029) &  13.013(0.132) & 108.112(0.887) \\ 
EQUAL & 0.067(0.002) & 0.361(0.009) &  1.264(0.028) &   4.892(0.105) &  20.622(0.521) \\ 
D-trace & 0.081(0.003) & 0.489(0.015) &  2.901(0.063) &  17.331(0.310) & 167.160(8.216) \\ 
EQUALs & 0.113(0.004) & 0.660(0.021) &  1.731(0.034) &   5.619(0.094) &  24.904(0.669) \\ 
		\hline \\[-1.8ex] 
		&\multicolumn{5}{c}{Case 3:  $\Omega=\Omega_3$}\\ \\[-1.8ex] 
		CLIME & 0.446(0.028) & 2.598(0.169) & 16.605(1.133) & 129.968(3.816) & 918.681(12.421) \\ 
glasso & 0.009(0.001) & 0.072(0.006) &  0.317(0.013) &   1.818(0.015) &   7.925(0.080) \\ 
BigQuic & 1.786(0.043) & 4.121(0.035) & 11.293(0.164) &  36.905(0.302) & 140.656(1.949) \\ 
glasso-ADMM & 0.517(0.051) & 1.182(0.027) &  3.516(0.079) &  14.467(0.128) & 102.048(4.502) \\ 
SCIO & 0.138(0.008) & 0.230(0.007) &  1.640(0.016) &  12.697(0.143) & 106.806(0.988) \\ 
EQUAL & 0.095(0.015) & 0.143(0.002) &  0.580(0.010) &   2.962(0.028) &  17.002(0.220) \\ 
D-trace & 0.057(0.025) & 0.164(0.003) &  1.253(0.048) &  10.758(0.258) & 131.724(5.031) \\ 
EQUALs & 0.039(0.011) & 0.226(0.003) &  0.768(0.014) &   3.430(0.026) &  19.133(0.224) \\ 
		\hline \\[-1.8ex] 
	\end{tabular} }
\end{table}

\subsection{Estimation accuracy}

The second simulation is designed to evaluate the performance of estimation accuracy. Given the true precision matrix $\Omega$ and an estimator $\hat{\Omega}$, we report the following four loss functions:
\begin{align*}
\mbox{loss1}&=\frac{1}{\sqrt{p}}\|\Omega-\hat{\Omega}\|_2,~~\mbox{loss2}=\|\Omega-\hat{\Omega}\|, \\
\mbox{loss3}&=\sqrt{\frac{1}{p}\{\tr(\Omega^{-1} \hat{\Omega})-\log|\Omega^{-1} \hat{\Omega}|-p \}},\\
\mbox{loss4}&=\sqrt{\frac{1}{p}\{\tr(\hat{\Omega}^\T \Omega^{-1} \hat{\Omega})/2-\tr(\hat{\Omega})+\tr(\Omega)/2 \}},
\end{align*}
where $\mbox{loss1}$ is the scaled Frobenius loss, $\mbox{loss2}$ is the spectral loss, $\mbox{loss3}$ is the normalized Stein's loss which is related to the Gaussian likelihood and $\mbox{loss4}$ is related to the quadratic loss.  

Table \ref{tab2} reports the simulation results based on 100 replications where the tuning parameter is chosen by five-fold cross-validations. We can see that the performance of all three estimators are comparable, indicating that the penalized quadratic loss estimators are also reliable for high dimensional precision matrix estimation. As shown in Table \ref{tab1} , the computation for quadratic loss estimator are much faster than glasso.  We also observe that the EQUALs estimator based on the symmetric loss \eqref{zou} has slightly smaller estimation error than EQUAL based on \eqref{liu}, which indicates that considering the symmetry structure does help improve the estimation accuracy.   Moreover, to check the singularity of the estimation, we report the minimum eigenvalue for each estimator in the final column of Table \ref{tab2}. We can see when the tuning parameter is suitably chosen, the penalized quadratic loss estimator is also positive definite.

\begin{table}[!htbp] 
	\caption{The estimation error (standard deviation) for the precision matrix estimation.}
	\label{tab2} 
		\resizebox{\textwidth}{!}{%
	\begin{tabular}{@{\extracolsep{5pt}} cccccc} 
		\\[-1.8ex]\hline 
		\hline \\[-1.8ex] 
		& loss1 & loss2 & loss3 & loss4 & min-Eigen \\ 
		\hline \\[-1.8ex] 
		&\multicolumn{5}{c}{Case 1: $\Omega=\Omega_1$}\\	\\[-1.8ex] 
				&\multicolumn{5}{c}{$p=500$}\\	\\[-1.8ex] 
		EQUAL & 0.707(0.005) & 2.028(0.016) & 0.329(0.003) & 0.344(0.003) & 0.364(0.011) \\ 
		EQUALs & 0.664(0.010) & 1.942(0.026) & 0.297(0.005) & 0.320(0.004) & 0.333(0.011) \\ 
		glasso & 0.685(0.006) & 1.973(0.015) & 0.313(0.002) & 0.332(0.001) & 0.216(0.010) \\ 
		&\multicolumn{5}{c}{$p=1000$}\\	\\[-1.8ex] 
		EQUAL & 0.701(0.008) & 2.033(0.020) & 0.331(0.004) & 0.344(0.004) & 0.361(0.012) \\ 
		EQUALs & 0.681(0.003) & 1.983(0.013) & 0.314(0.002) & 0.331(0.002) & 0.353(0.011) \\ 
		glasso & 0.690(0.005) & 1.984(0.012) & 0.335(0.004) & 0.344(0.002) & 0.172(0.012) \\ 
		&\multicolumn{5}{c}{$p=2000$}\\	\\[-1.8ex] 
		EQUAL & 0.860(0.106) & 2.351(0.190) & 0.446(0.066) & 0.426(0.049) & 0.322(0.023) \\ 
		EQUALs & 0.666(0.020) & 1.984(0.042) & 0.317(0.008) & 0.331(0.007) & 0.348(0.011) \\ 
		glasso & 0.695(0.004) & 1.992(0.012) & 0.365(0.007) & 0.361(0.003) & 0.118(0.011) \\ 
	\hline \\[-1.8ex] 
		&\multicolumn{5}{c}{Case 2: $\Omega=\Omega_2$} \\  \\[-1.8ex] 
		&\multicolumn{5}{c}{$p=500$}\\	\\[-1.8ex] 
EQUAL & 0.508(0.010) & 1.178(0.045) & 0.273(0.004) & 0.286(0.004) & 0.555(0.018) \\ 
EQUALs & 0.465(0.010) & 1.116(0.045) & 0.240(0.003) & 0.254(0.004) & 0.500(0.015) \\ 
glasso & 0.530(0.011) & 1.179(0.037) & 0.234(0.003) & 0.269(0.003) & 0.267(0.013) \\ 
&\multicolumn{5}{c}{$p=1000$}\\	\\[-1.8ex] 
EQUAL & 0.605(0.011) & 1.323(0.036) & 0.304(0.005) & 0.326(0.005) & 0.578(0.017) \\ 
EQUALs & 0.550(0.008) & 1.272(0.039) & 0.267(0.003) & 0.289(0.004) & 0.527(0.015) \\ 
glasso & 0.542(0.008) & 1.217(0.026) & 0.260(0.005) & 0.289(0.002) & 0.211(0.015) \\ 
&\multicolumn{5}{c}{$p=2000$}\\	\\[-1.8ex] 
EQUAL & 0.555(0.006) & 1.294(0.051) & 0.291(0.003) & 0.307(0.003) & 0.560(0.015) \\ 
EQUALs & 0.539(0.008) & 1.253(0.043) & 0.279(0.003) & 0.294(0.003) & 0.550(0.014) \\ 
glasso & 0.558(0.005) & 1.263(0.019) & 0.297(0.006) & 0.317(0.004) & 0.144(0.011) \\ 
		\hline \\[-1.8ex] 
				&\multicolumn{5}{c}{Case 3: $\Omega=\Omega_3$} \\  \\[-1.8ex] 
		&\multicolumn{5}{c}{$p=500$}\\	\\[-1.8ex] 
		EQUAL & 1.181(0.013) & 4.336(0.221) & 0.427(0.000) & 0.451(0.000) & 0.110(0.011) \\ 
EQUALs & 1.183(0.014) & 4.342(0.221) & 0.427(0.001) & 0.451(0.000) & 0.126(0.011) \\ 
glasso & 1.188(0.013) & 4.351(0.218) & 0.420(0.001) & 0.450(0.000) & 0.087(0.009) \\ 
		&\multicolumn{5}{c}{$p=1000$}\\	\\[-1.8ex] 
		EQUAL & 1.183(0.009) & 4.276(0.143) & 0.428(0.000) & 0.453(0.001) & 0.101(0.008) \\ 
EQUALs & 1.189(0.009) & 4.337(0.148) & 0.426(0.000) & 0.451(0.000) & 0.106(0.007) \\ 
glasso & 1.189(0.010) & 4.349(0.153) & 0.421(0.001) & 0.450(0.000) & 0.080(0.006) \\ 
		&\multicolumn{5}{c}{$p=2000$}\\	\\[-1.8ex] 
	EQUAL & 1.217(0.009) & 4.649(0.120) & 0.436(0.001) & 0.458(0.001) & 0.097(0.006) \\ 
EQUALs & 1.239(0.006) & 4.563(0.102) & 0.450(0.002) & 0.461(0.001) & 0.062(0.006) \\ 
glasso & 1.190(0.007) & 4.399(0.109) & 0.422(0.001) & 0.450(0.000) & 0.076(0.004) \\ 
		\hline \\[-1.8ex] 
	\end{tabular} }
\end{table} 

\subsection{Local linear approximation}
\begin{figure}[!ht]
	\centerline{
		\begin{tabular}{ccc}
			\psfig{figure=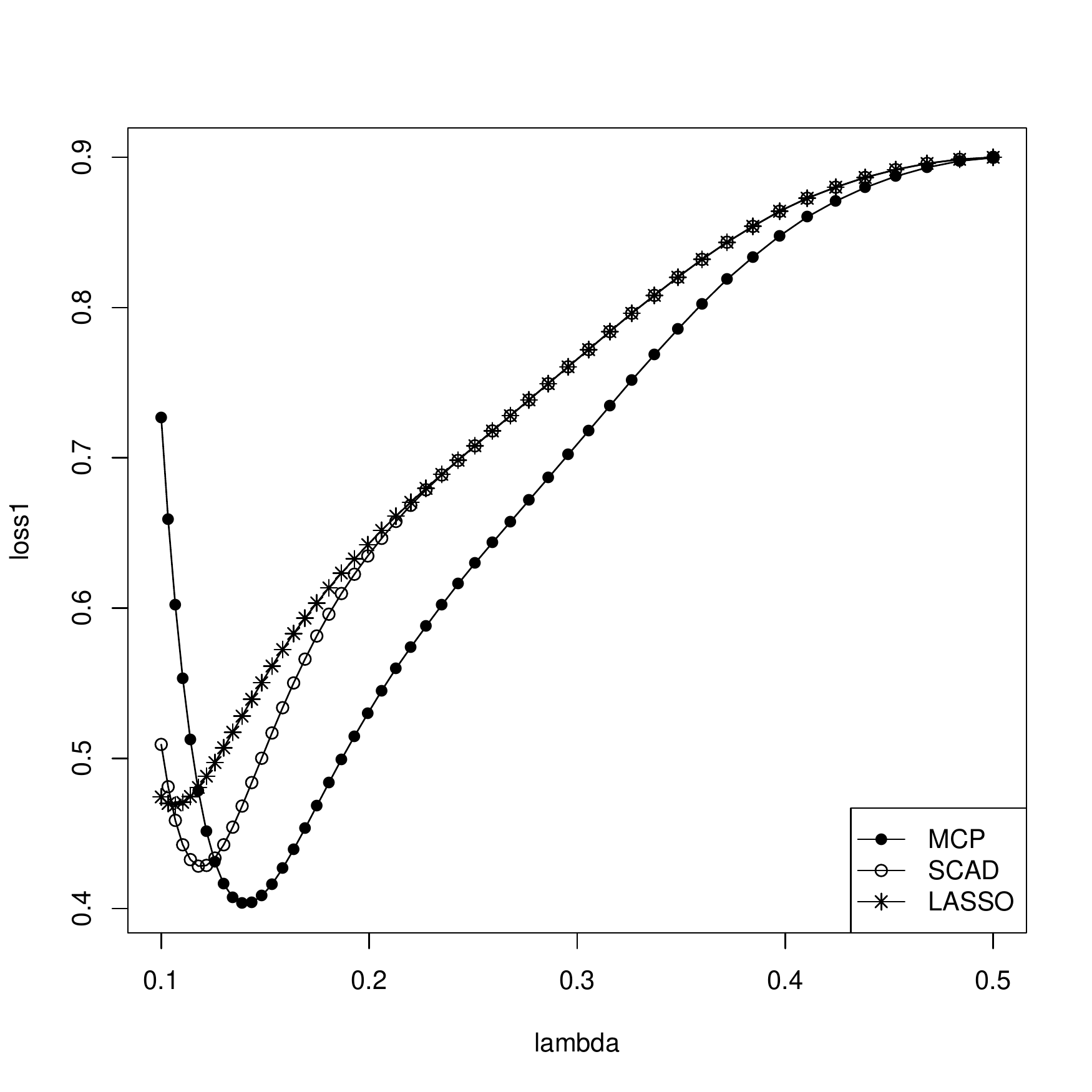,width=1.5 in,height=1.5in,angle=0} &
			\psfig{figure=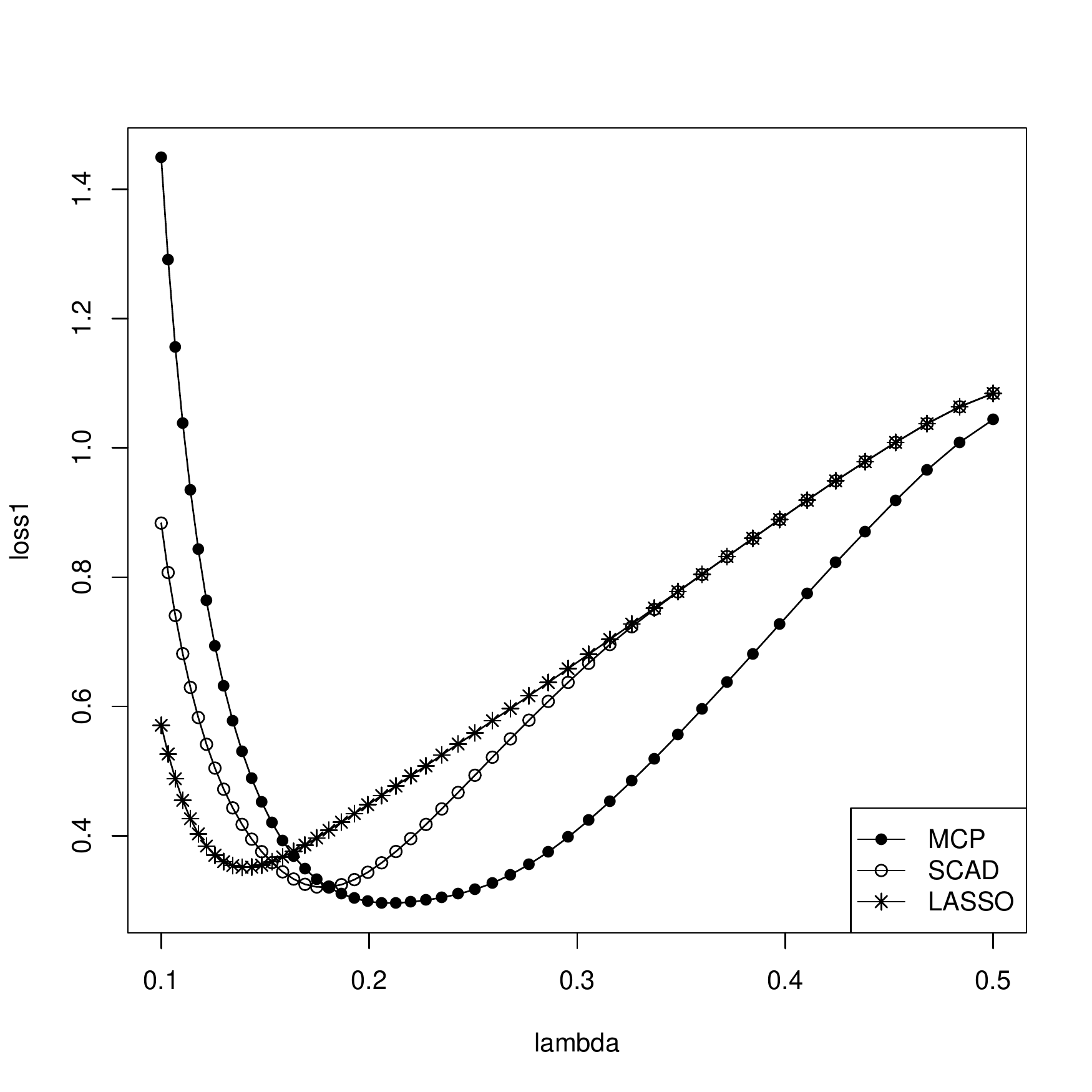,width=1.5 in,height=1.5in,angle=0} &
			\psfig{figure=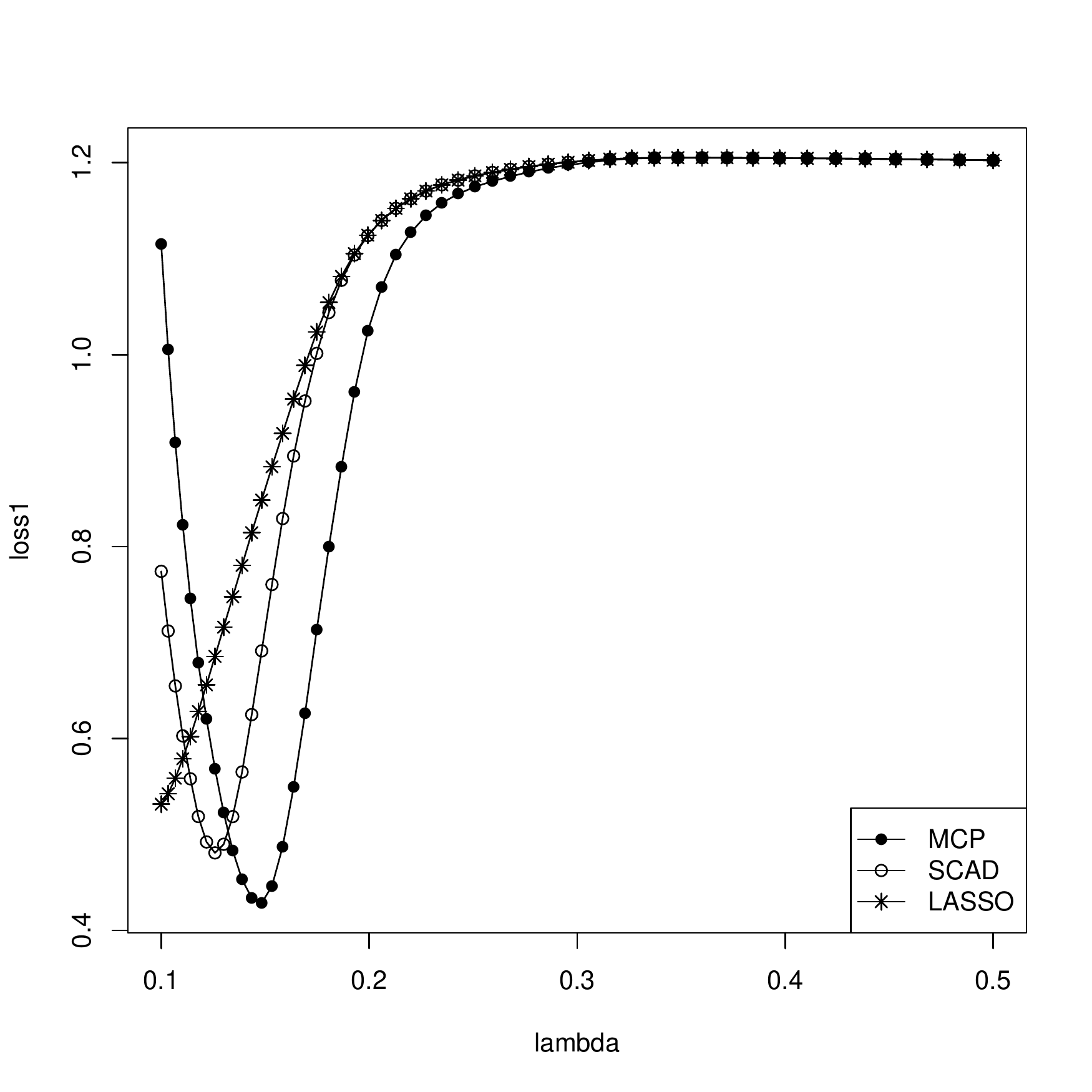,width=1.5 in,height=1.5in,angle=0}  \\
				Case 1: loss1 for $\Omega_1$ & Case 2: loss1 for $\Omega_2$ & Case 3: loss1 for $\Omega_3$ \\
						\psfig{figure=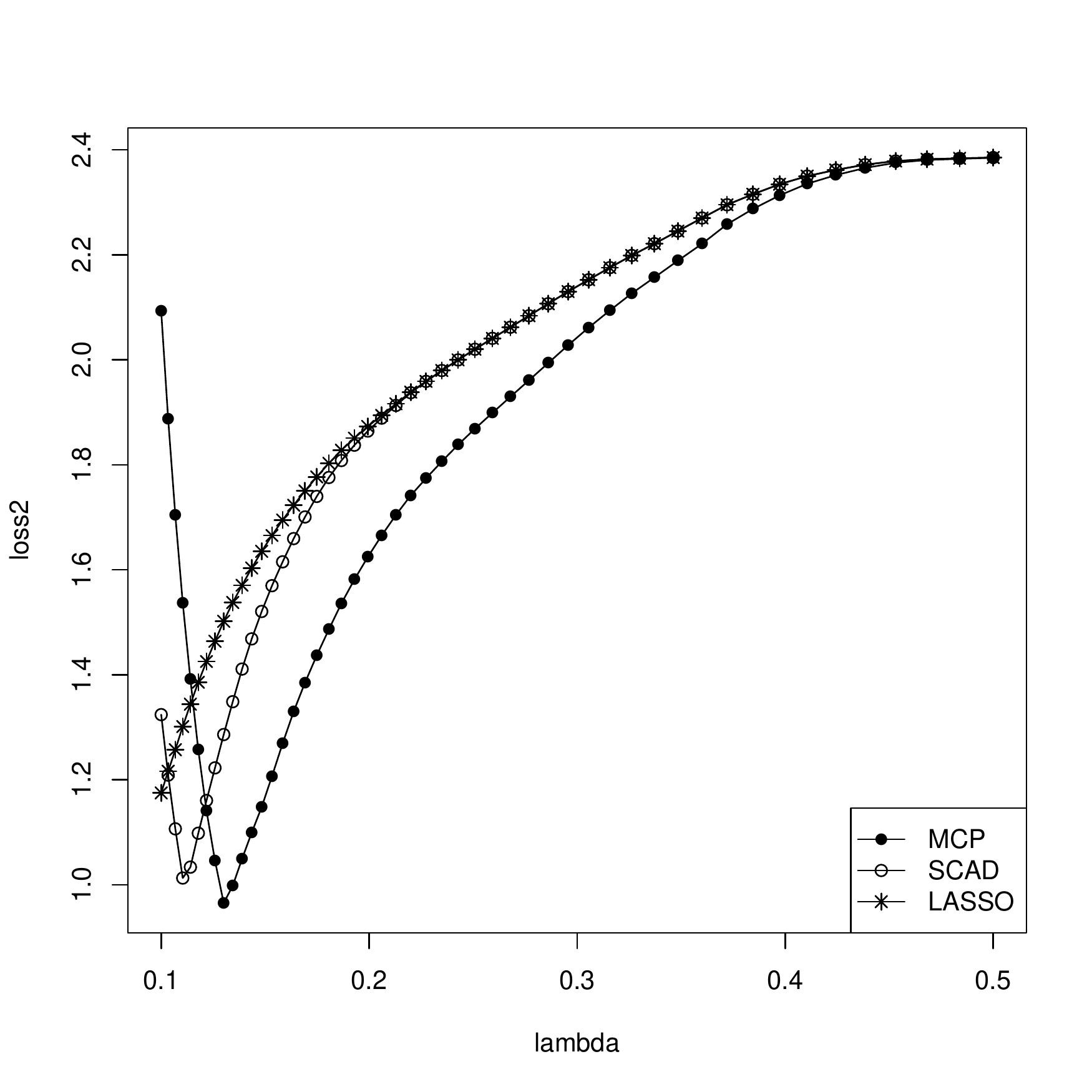,width=1.5 in,height=1.5in,angle=0} &
			\psfig{figure=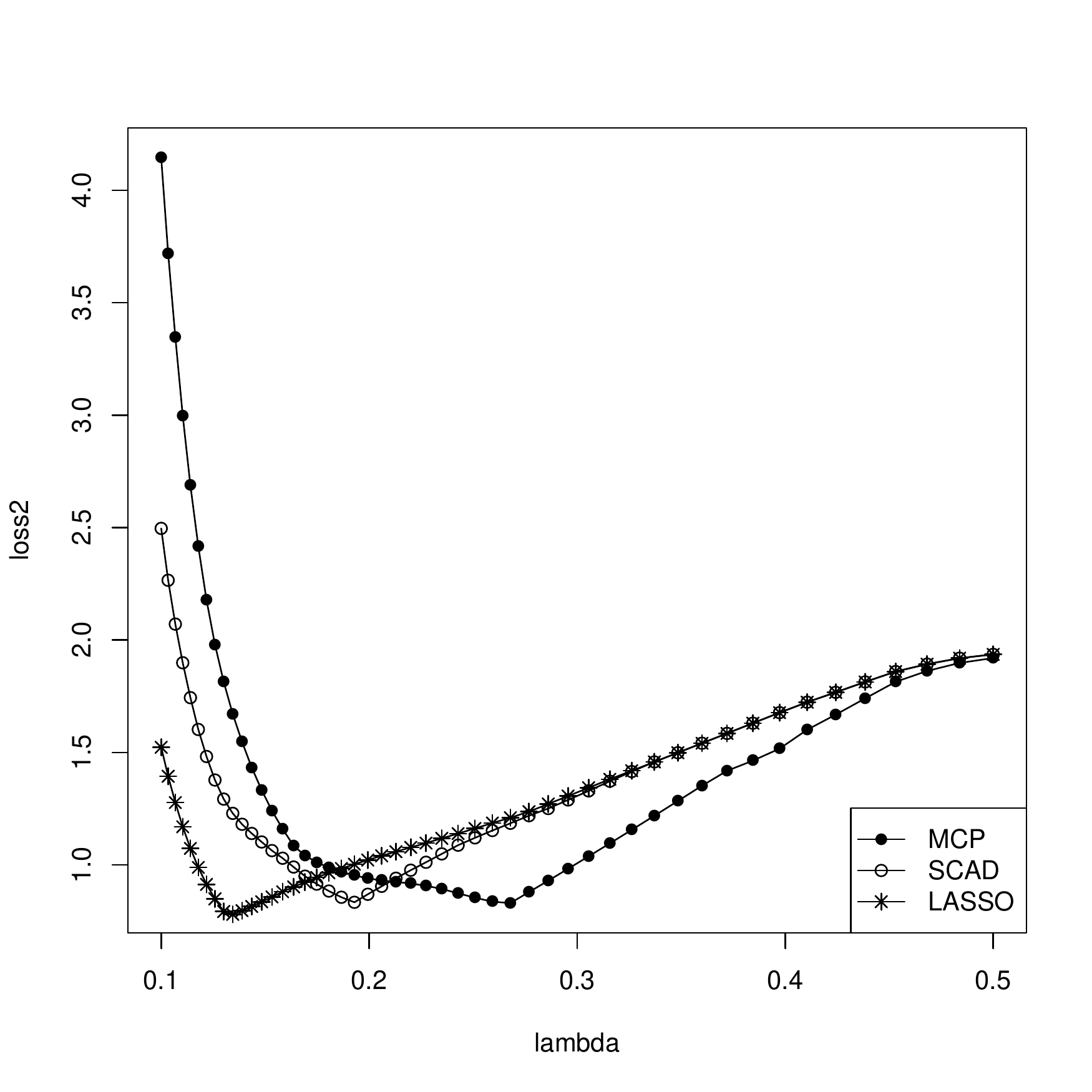,width=1.5 in,height=1.5in,angle=0} &
			\psfig{figure=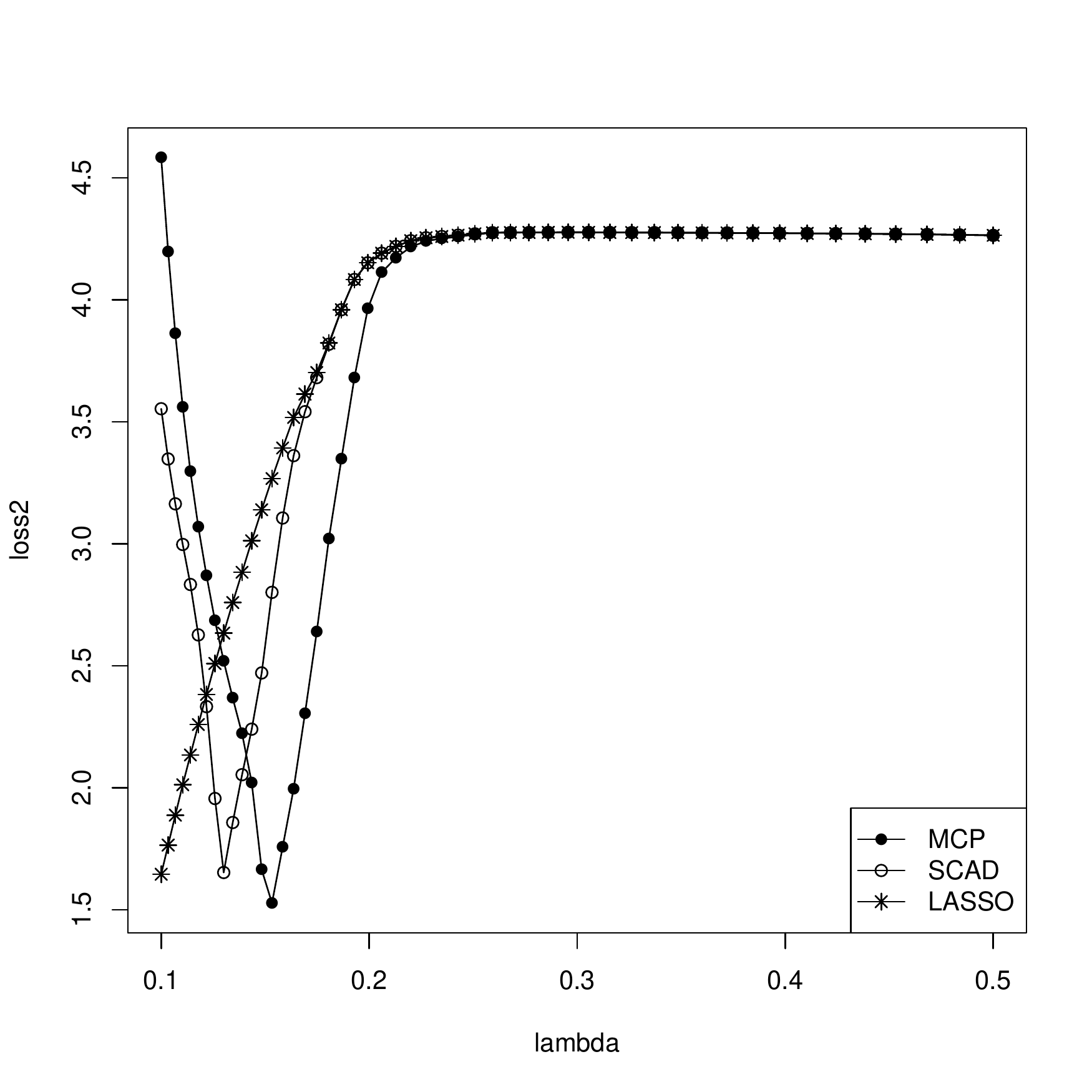,width=1.5 in,height=1.5in,angle=0}  \\
		Case 1: loss2 for $\Omega_1$ & Case 2: loss2 for $\Omega_2$ & Case 3: loss2 for $\Omega_3$ \\			
						\psfig{figure=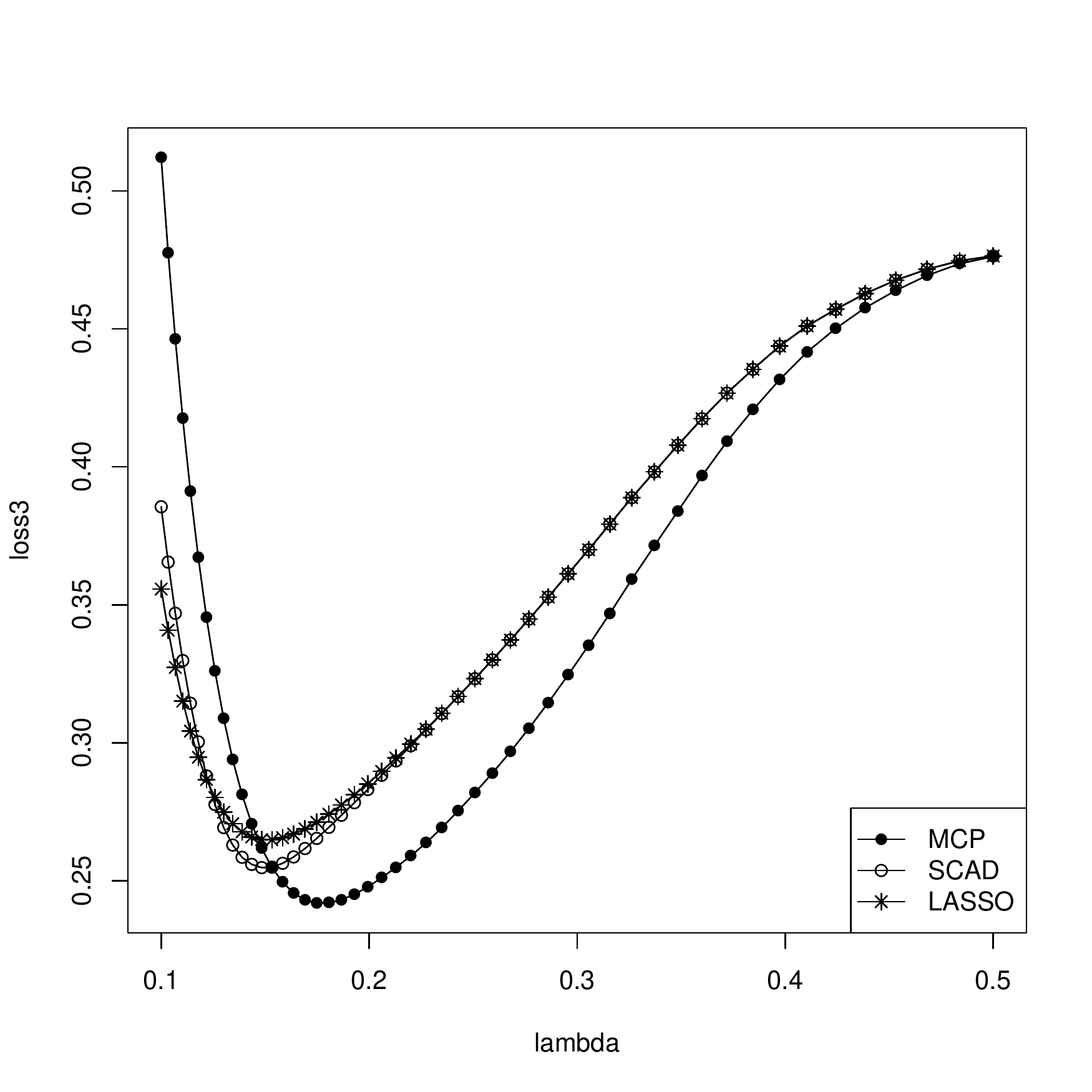,width=1.5 in,height=1.5in,angle=0} &
			\psfig{figure=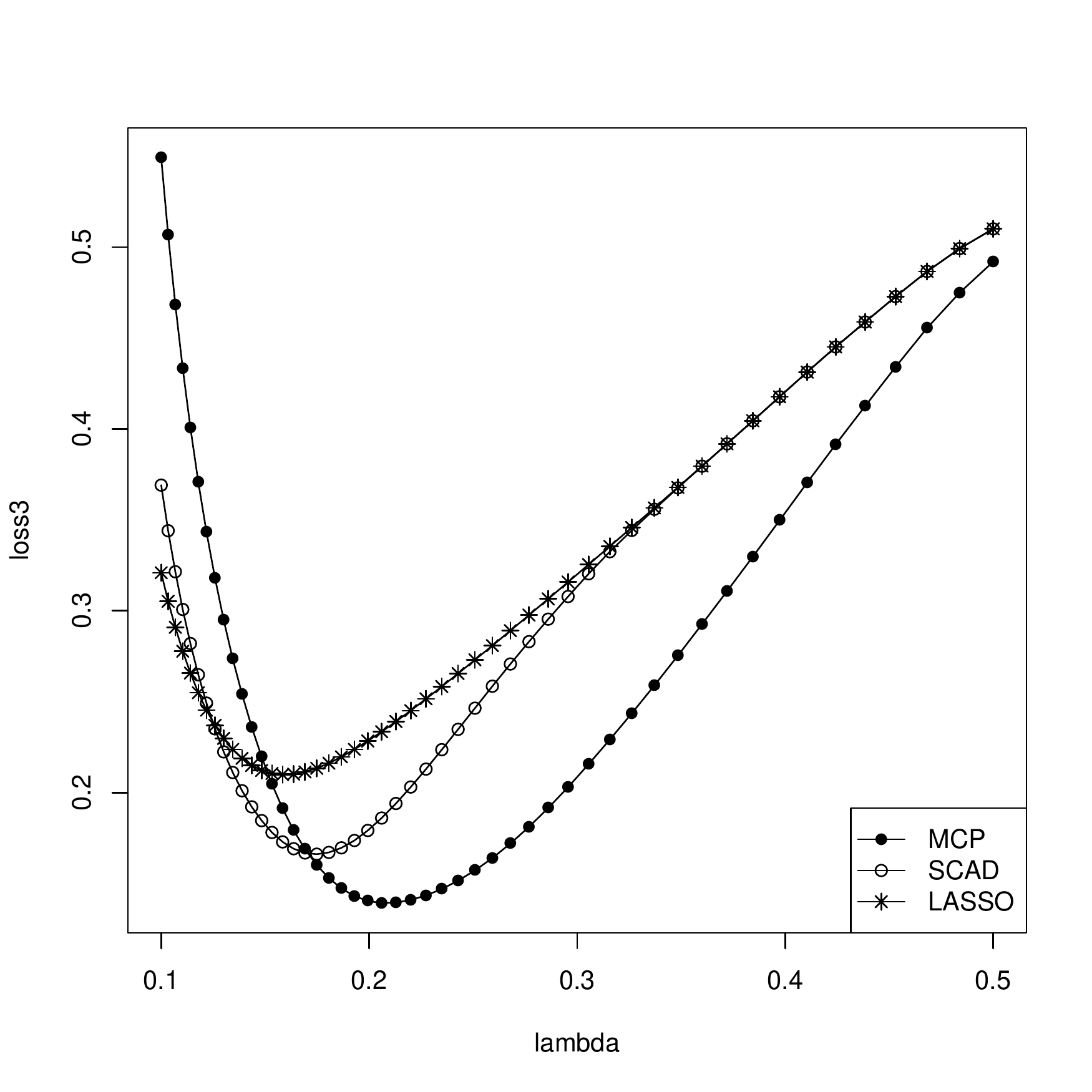,width=1.5 in,height=1.5in,angle=0} &
			\psfig{figure=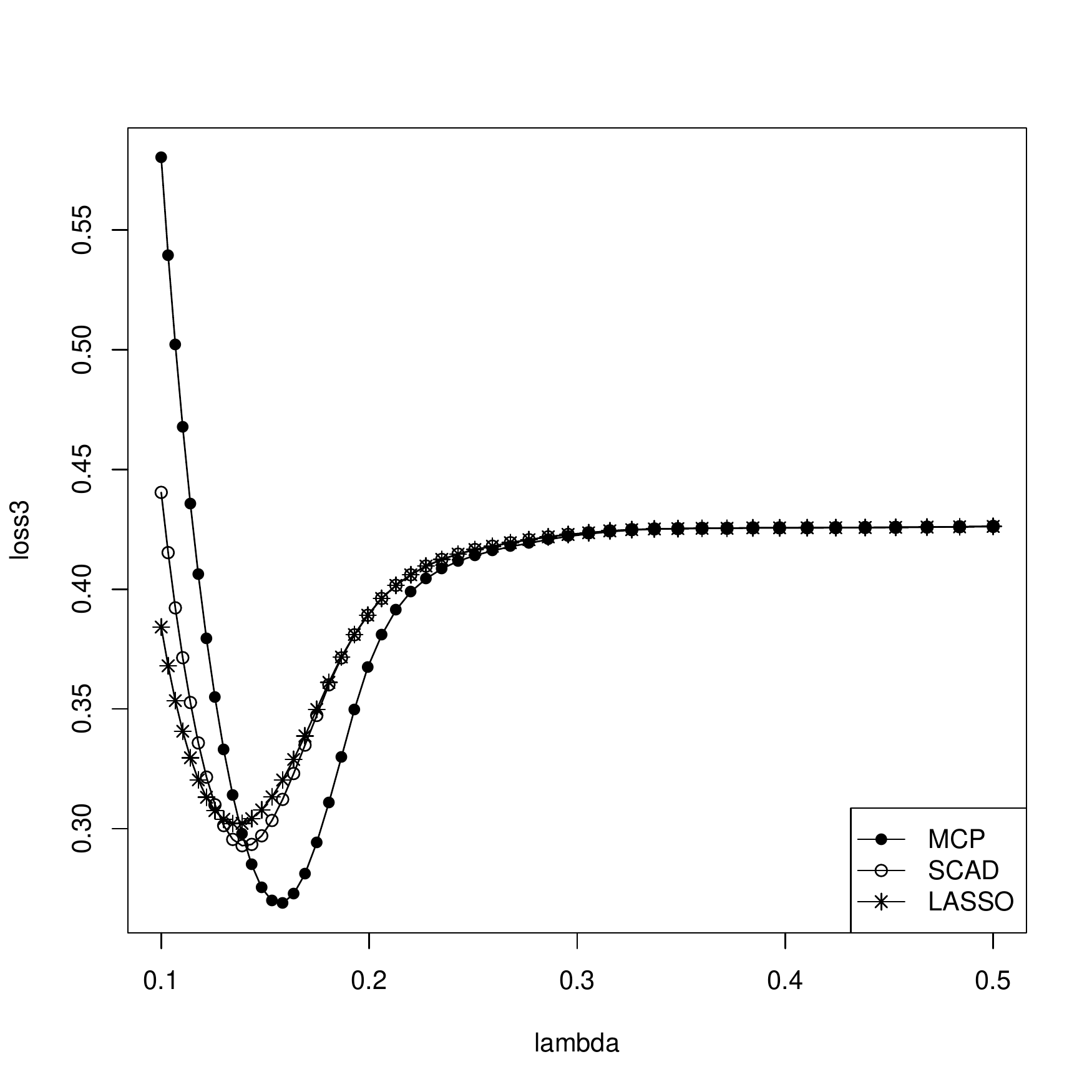,width=1.5 in,height=1.5in,angle=0}  \\
		Case 1: loss3 for $\Omega_1$ & Case 2: loss3 for $\Omega_2$ & Case 3: loss3 for $\Omega_3$ \\			
						\psfig{figure=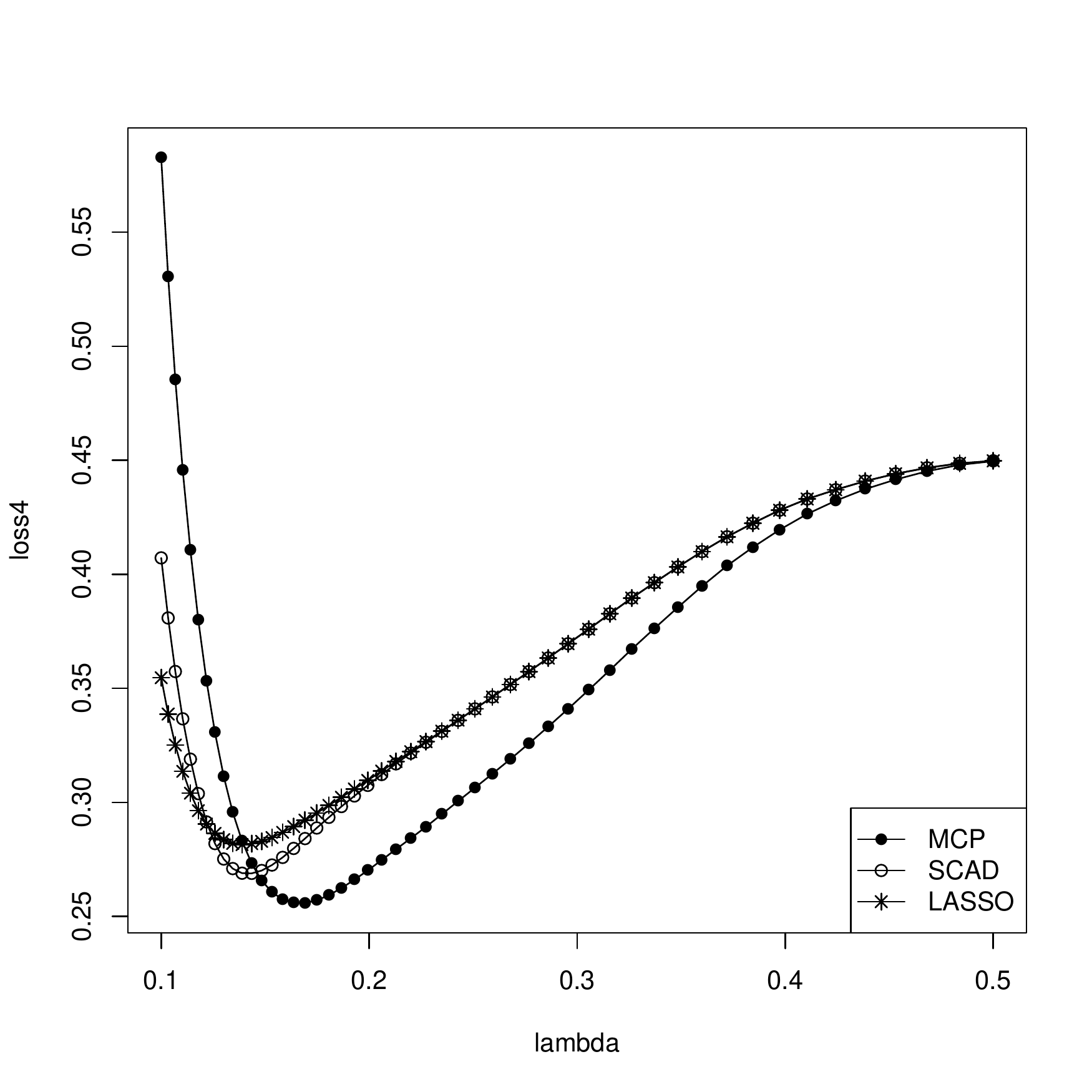,width=1.5 in,height=1.5in,angle=0} &
			\psfig{figure=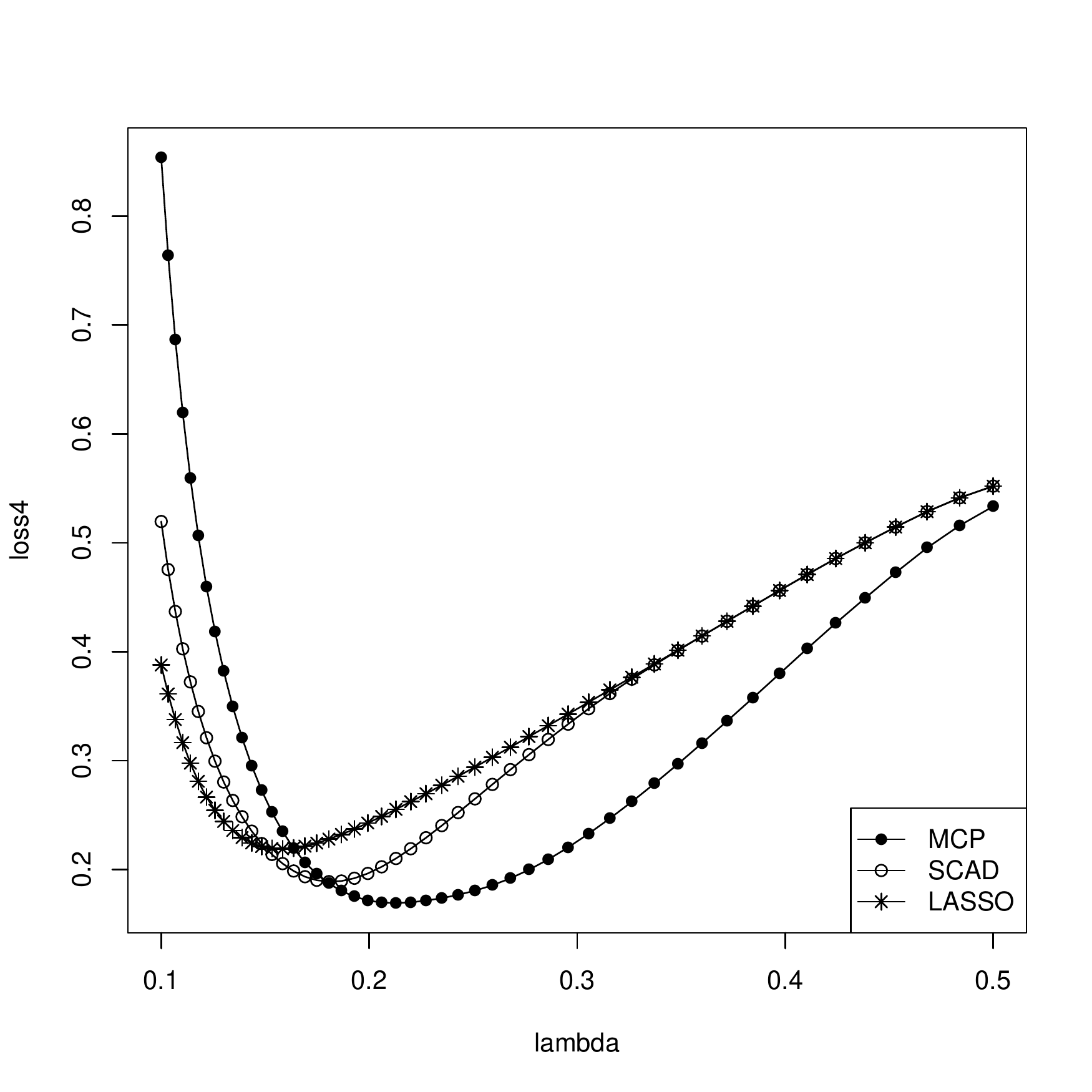,width=1.5 in,height=1.5in,angle=0} &
			\psfig{figure=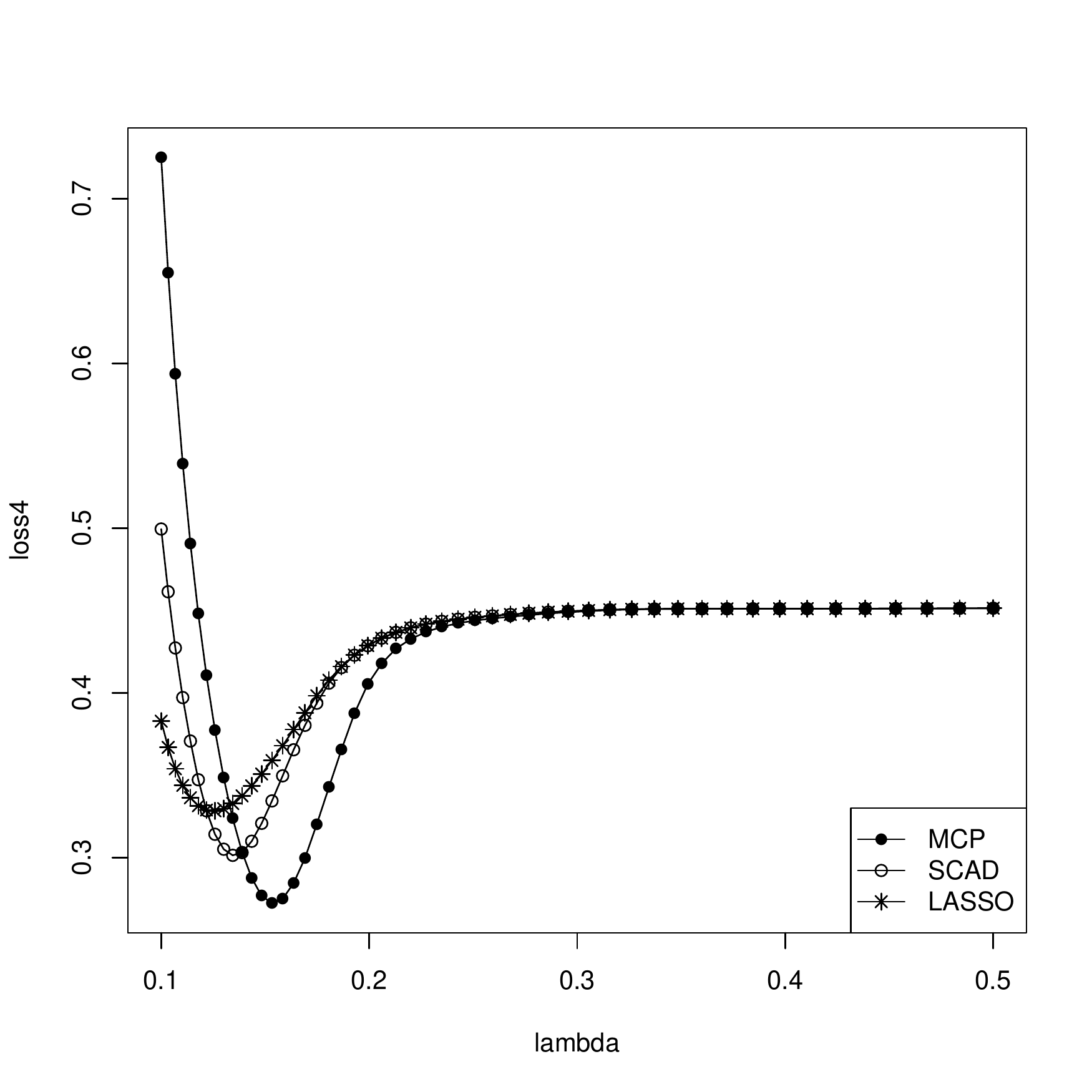,width=1.5 in,height=1.5in,angle=0}  \\
					Case 1: loss4 for $\Omega_1$ & Case 2: loss4 for $\Omega_2$ & Case 3: loss4 for $\Omega_3$ \\
	\end{tabular}
	}
	\caption{The performance of the quadratic loss based estimators with LASSO, SCAD and MCP penalties.}
	\label{fig1}
\end{figure}

In this part, we consider the estimator with more general penalized functions based on the one step local linear approximation proposed by \cite{zou2008one}. In details, we consider the SCAD penalty \citep{fan2001variable}: 
\begin{align*}
p_\lambda(x)=&\lambda |x| I(|x| \leq \lambda)+\frac{ \tau \lambda |x|-(x^2+\lambda^2)/2}{\tau-1}I(\lambda<|x|\leq \tau \lambda)\\
&+\frac{\lambda^2(\tau+1)}{2} I(|x|>\tau \lambda),~\tau=3.7,
\end{align*}
and MCP \citep{zhang2010nearly}:
\begin{align*}
p_\lambda(x)=\Big[\lambda |x|-\frac{x^2}{2 \tau}\Big] I(|x| \leq \tau \lambda)+\frac{\lambda^2 \tau}{2} I(|x|>\tau \lambda),~\tau=2.
\end{align*}
The new estimator is defined as 
\begin{align*}
\argmin_{\tiny {\Omega \in \mR^{p\times p}}}  L(\Omega)+\sum_{i \neq j} p_\lambda(\Omega_{ij}),
\end{align*}
where $L(\Omega)$ is the quadratic loss function and $p_\lambda(\cdot)$ is a penalty function. Following \citep{zou2008one}, we then seek to solve the following local linear approximation:
\begin{align*}
\argmin_{\tiny {\Omega \in \mR^{p\times p}}}  L(\Omega)+\sum_{i \neq j} p'_\lambda(\Omega^{(0)}_{ij}) |\Omega_{ij}|,
\end{align*}
where $\Omega^{(0)}$ is an initial estimator. We consider Cases 1-3 with $n=p=200$. For each tuning parameter $\lambda$, we calculate the LASSO solution $\hat{\Omega}_{\lambda}$, which is set to be the initial estimator, and calculate the one-step estimator for the MCP penalty and SCAD penalty respectively.  Figure \ref{fig1} reports the four loss functions defined above based on the LASSO, SCAD and MCP penalties, respectively.  For brevity, we only report the estimation for EQUALs. From Figure \ref{fig1}, we can see that SCAD and MCP penalties do produce slightly better estimation results.

\subsection{Real data analysis }
\begin{figure}[!ht]\small
	\centerline{
		\begin{tabular}{cc}				
			\psfig{figure=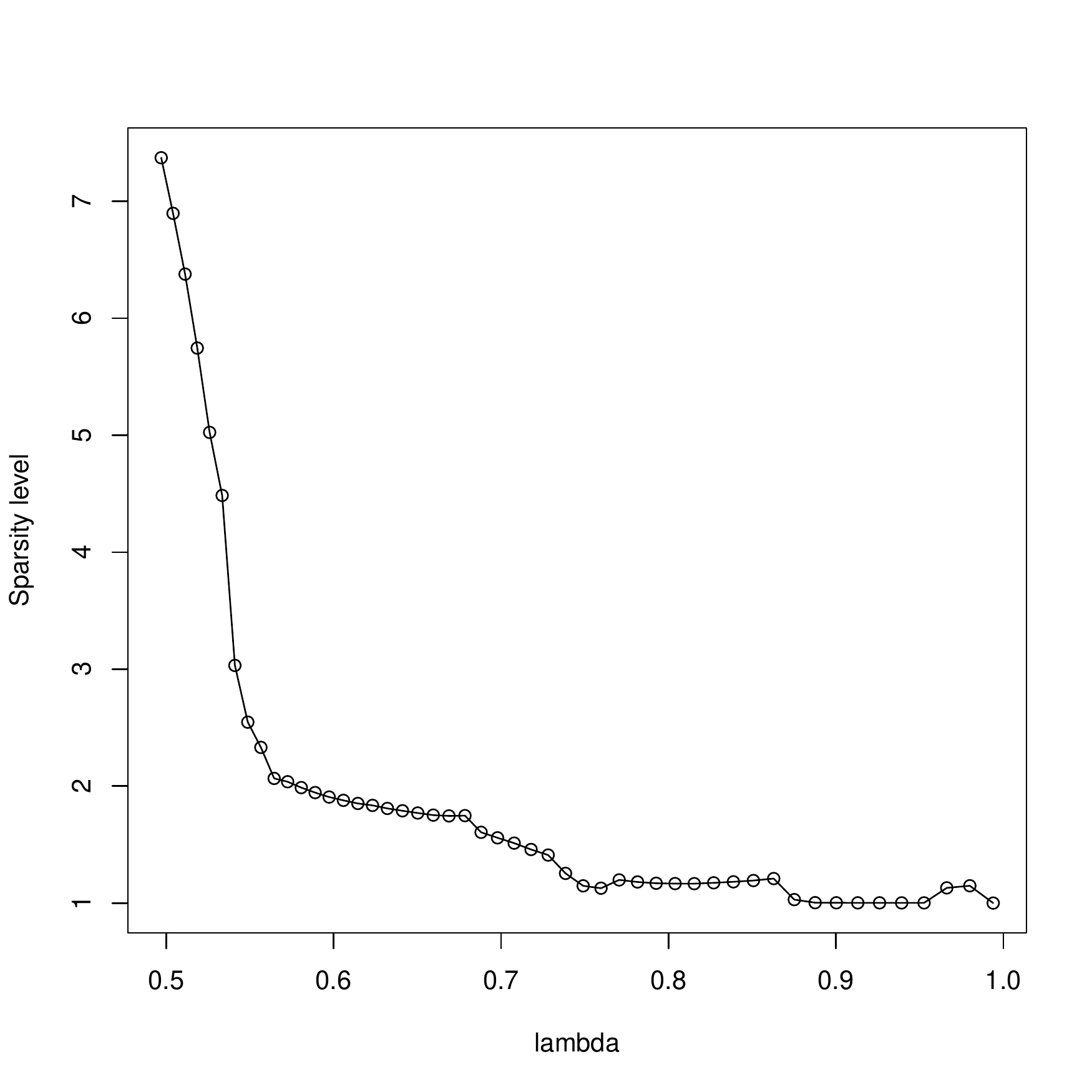,width=2.5 in,height=2.5 in,angle=0} &
			\psfig{figure=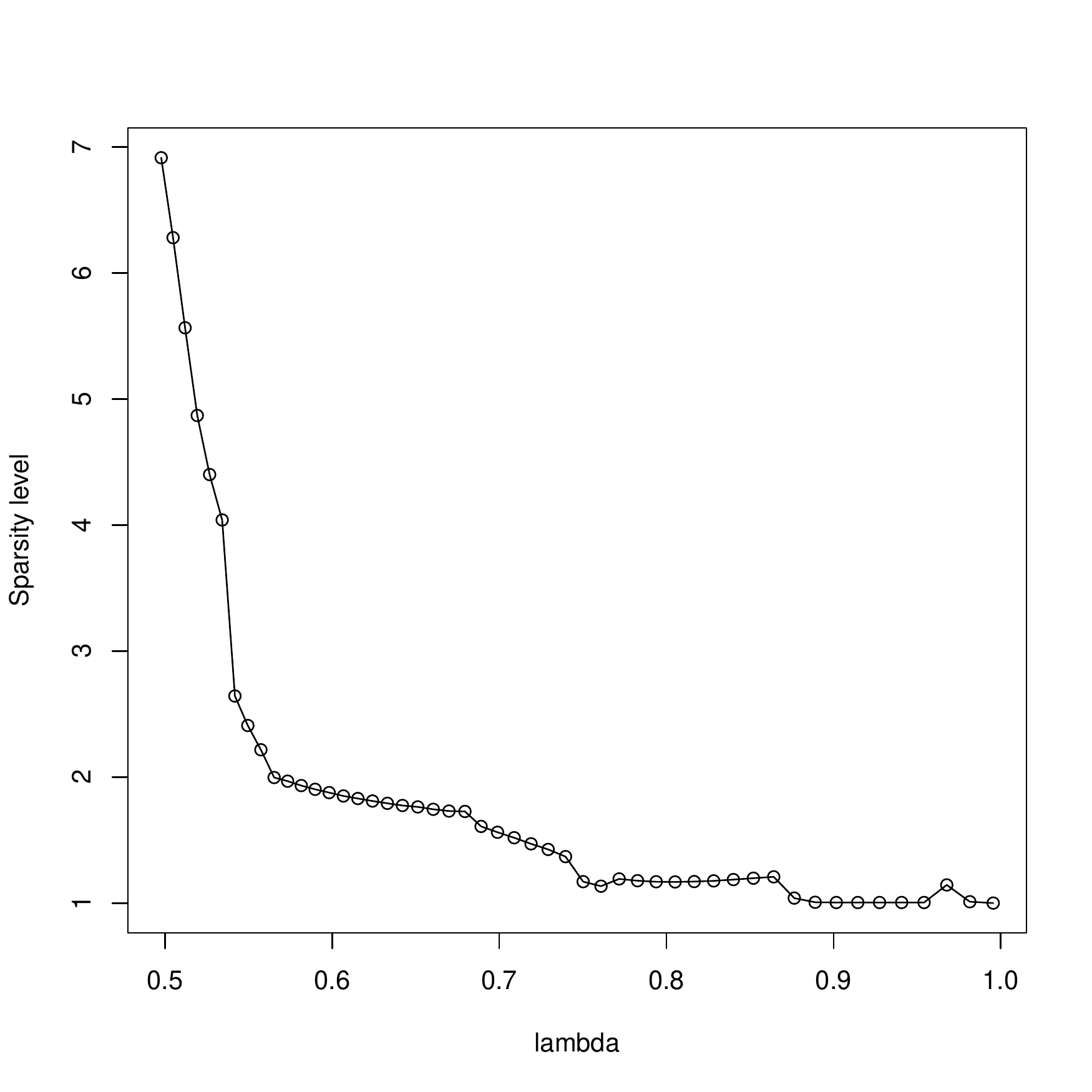,width=2.5 in,height=2.5 in,angle=0} \\
			(a) Solution path for control subjects
			& (b) Solution path for prostate cancer subjects \\
			\psfig{figure=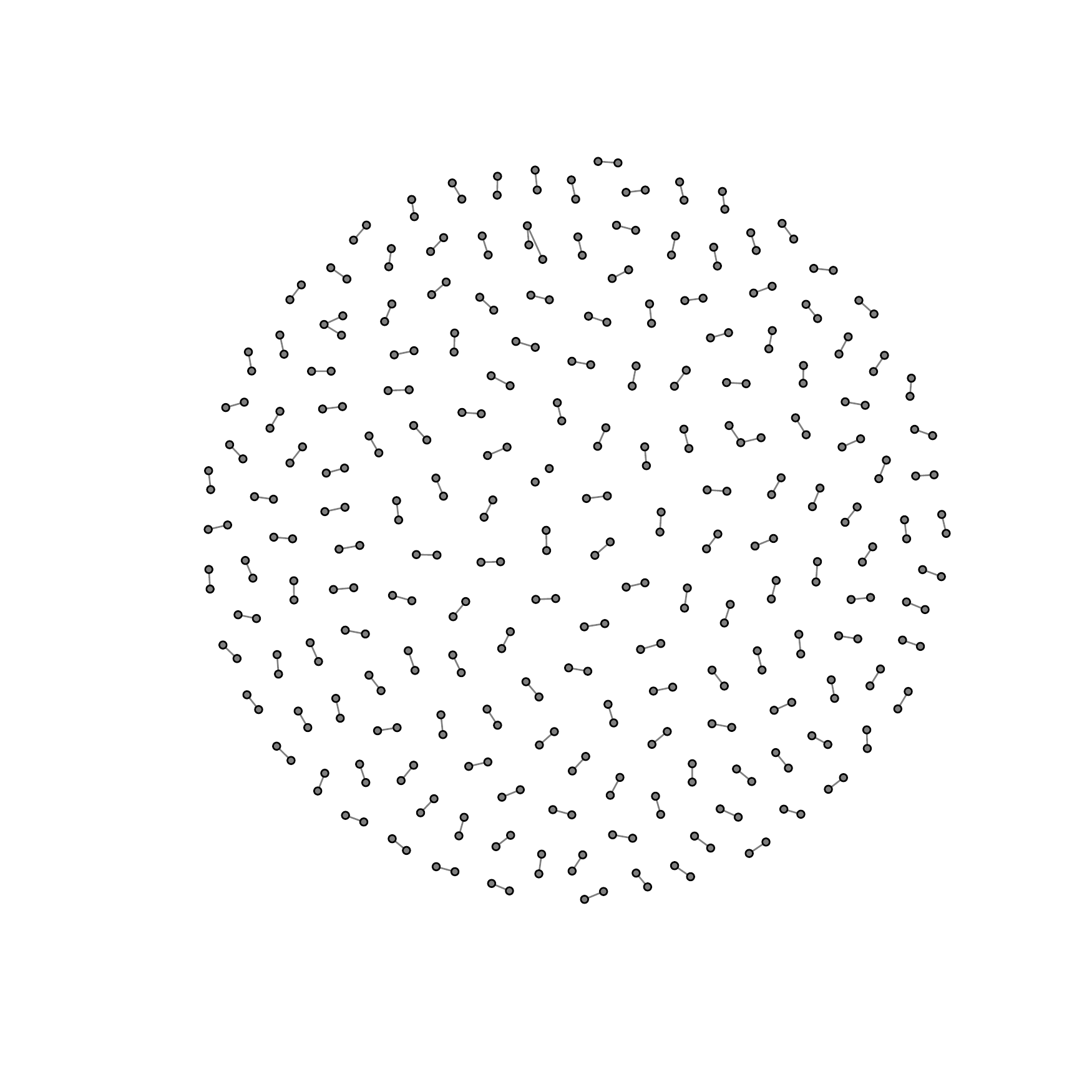,width=2.5 in,height=2.5 in,angle=0} &
			\psfig{figure=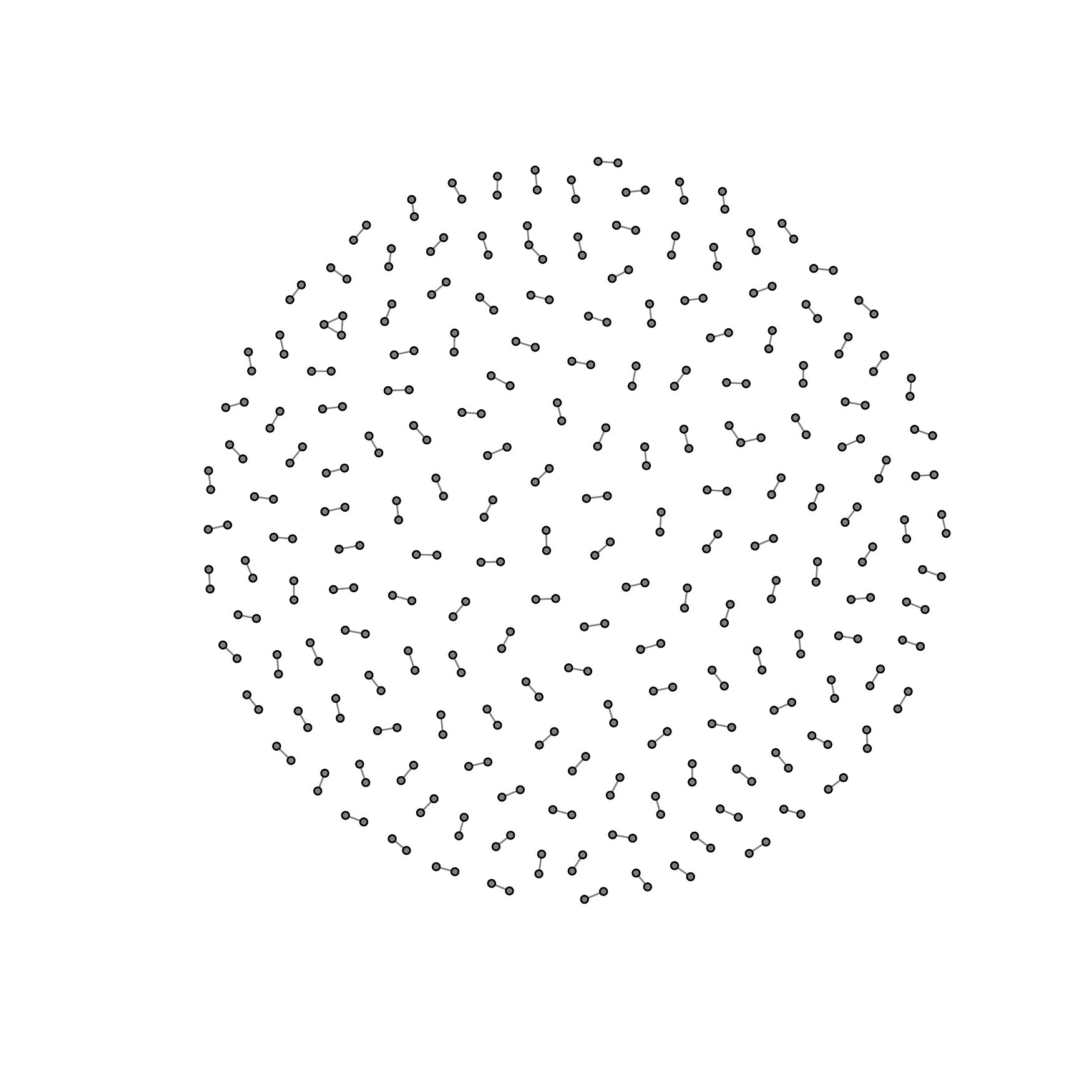,width=2.5 in,height=2.5 in,angle=0} \\
			(c) Estimated networks for control subjects 
			& (d) Estimated networks for prostate cancer subjects \\
		\end{tabular}
	}
	\caption{Estimation for the Prostate data using EQUALs. Upper panel: sparsity level (average number of non-zero elements for each row/column) versus $\lambda$. Lower panel: network graphs for the two patient groups when $\lambda =0.75$. }
	\label{real1}
\end{figure}

Finally, we apply our proposal to two real data. The first one is the Prostate dataset which is publicly  available at \url{https://web.stanford.edu/~hastie/CASI_files/DATA/prostate.html}. The data records 6033 genetic activity measurements for the control group (50 subjects) and the prostate cancer group (52 subjects). Here, the data dimension $p$ is 6033 and the sample size $n$ is 50 or 52. We estimate the $6033 \times 6033$ precision for each group. Since our EQUAL and EQUALs give similar results, we only report the estimation for EQUALs. It took less than 20 minutes for EQUALs to obtain the solution paths while ``glasso" cannot produce the solution due to out of memory in R. The sparsity level of the solution paths are plotted in the upper panel of Figure  \ref{real1}. To compare the precision matrices between the two groups, the network graphs of the EQUALs estimators with tuning $\lambda=0.75$  are provided in the lower panel of Figure  \ref{real1}.

\begin{figure}[!ht]\small
	\centerline{
		\begin{tabular}{cc}				
			\psfig{figure=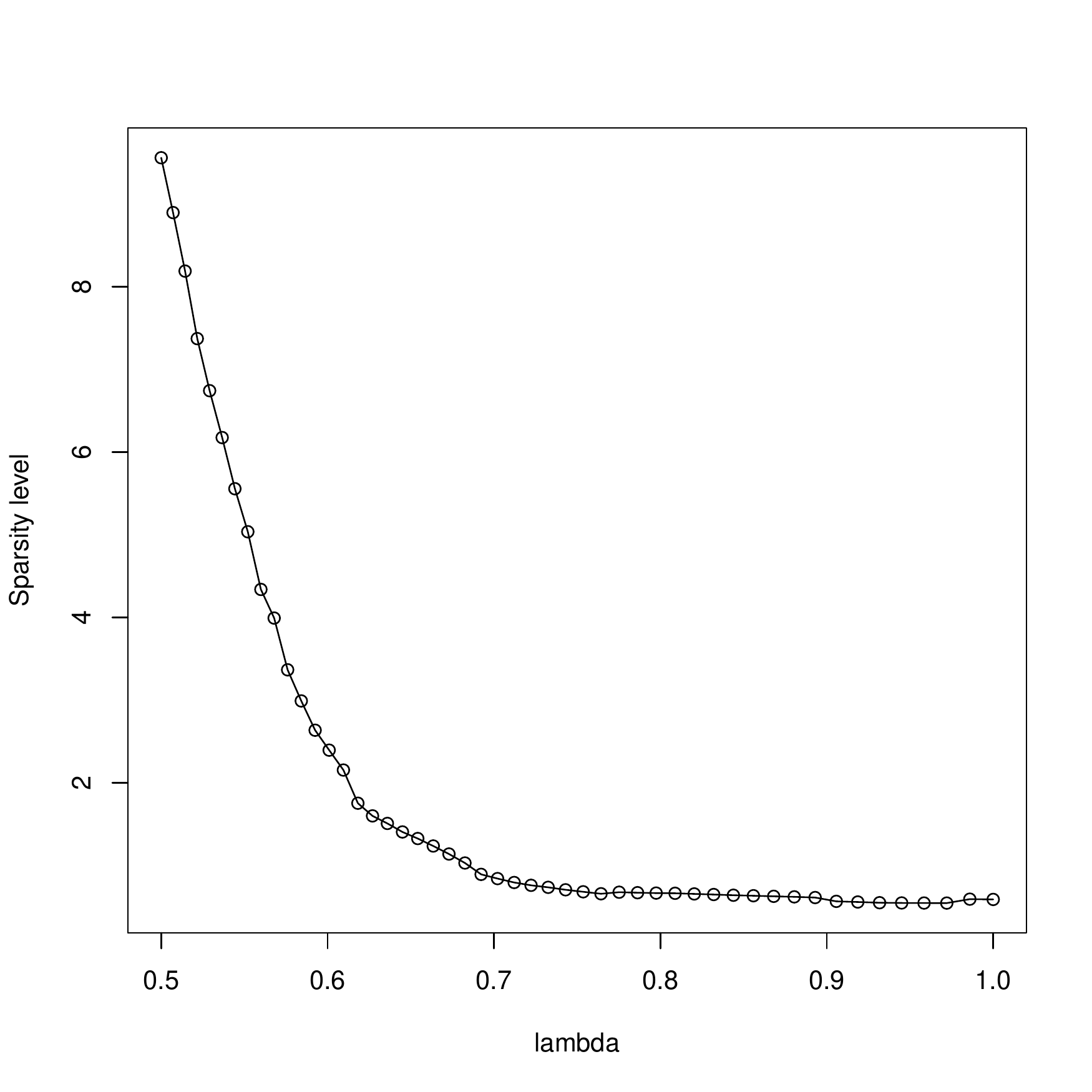,width=2.5 in,height=2.5 in,angle=0} &
			\psfig{figure=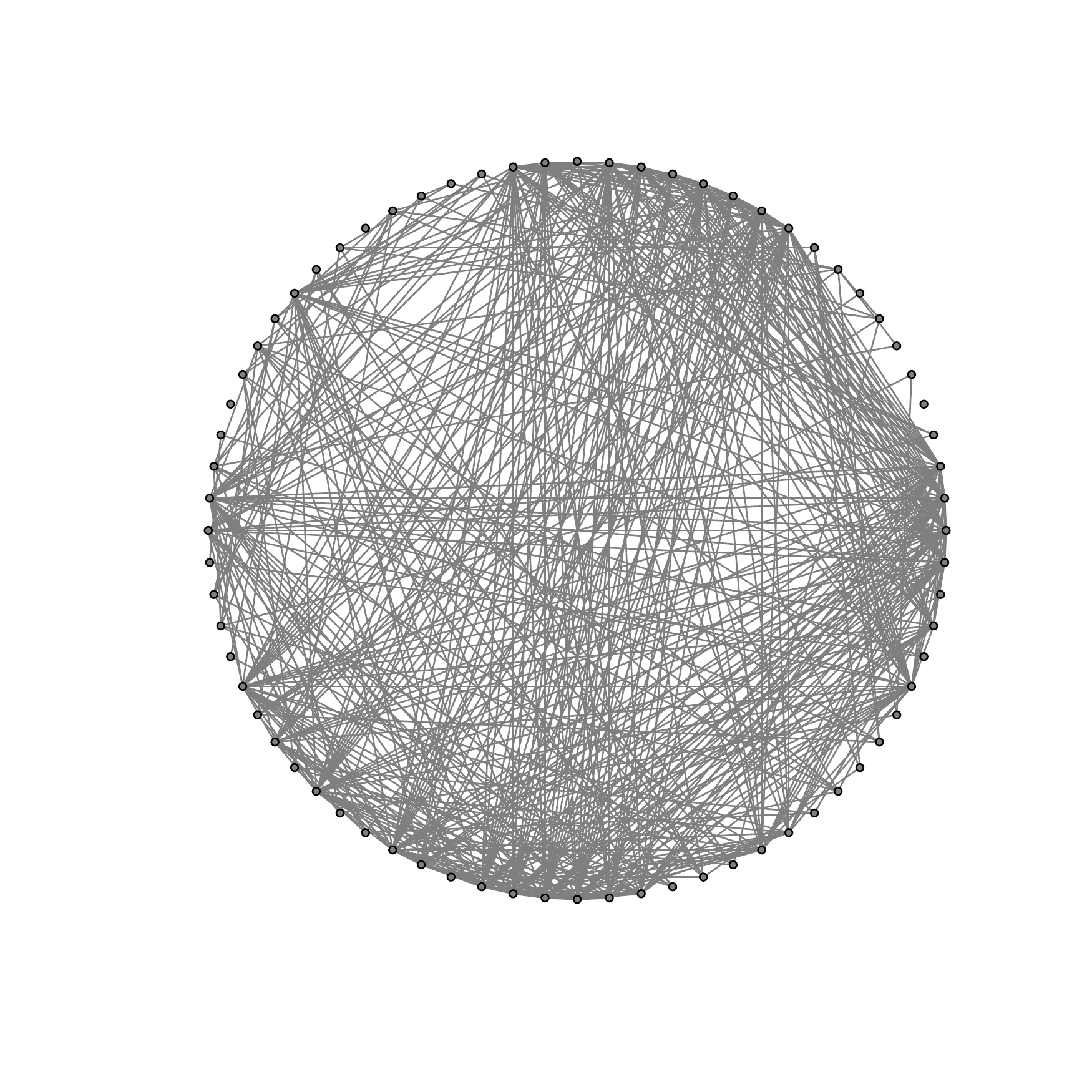,width=2.5 in,height=2.5 in,angle=0} \\
			(a) Solution path & (b) Estimated network with top 1\% links \\
		\end{tabular}
	}
	\caption{Estimation for the Leukemia data using EQUALs. (a) Sparsity level (average number of non-zero elements for each row/column) versus $\lambda$.  (b) Network graphs when $\lambda =0.6095$. }
	\label{real2}
\end{figure}

The second dataset is the leukemia data, which is publicly available at  \url{http://web.stanford.edu/~hastie/CASI_files/DATA/leukemia_big.csv.}
The dataset consists of 7128 genes for 47 acute lymphoblastic leukemia (ALL) patients. It took about 45 minutes for EQUALs to obtain the solution path and again,  ``glasso" fails to produce the results due to the vast memory requirement issue in R. The solution path and the network for top 1\% nodes with most links when $\lambda=0.6095$ are presented in Figure \ref{real2}.  

\section*{Acknowledgments}
We thank the Editor, an Associate Editor, and two anonymous reviewers for their insightful comments. Wang is partially supported by the Shanghai Sailing Program 16YF1405700, National Natural Science Foundation of China 11701367 and 11825104. Jiang is partially supported by the Early Career Scheme from Hong Kong Research Grants Council PolyU 253023/16P, and internal Grants PolyU 153038/17P.
\newpage
\section*{Appendix}
Throughout the proofs, we will use two important results of the Kronecker product 
	\begin{align*}
	\vec(AXB)=&(B^\T \otimes A ) \vec(X),~~(A \otimes B)(C \otimes D)=(AC) \otimes (BD),
	\end{align*}
where $X, A, B, C$ and $D$ are matrices of such size that one can form the matrix products.
\subsection{Proofs of Proposition \ref{prop0}}
The main techniques used for the proofs is the well-known Woodbury matrix identity. In details, \eqref{f1} is the direct application of the Woodbury matrix identity and \eqref{f2} can be obtained by invoking the identity repeatedly. The derivation involves lengthy and tedious calculations. Here, we simply prove the proposition by verifying the results. 

For the first formula \eqref{f1}, we have
\begin{align*}
& (S+\rho I_p)(\rho^{-1} I_p-\rho^{-1} U \Lambda_1 U ^\T)\\
=& (U \Lambda U ^\T+\rho I_p)(\rho^{-1} I_p-\rho^{-1} U \Lambda_1 U ^\T)\\
=& \rho^{-1} U \Lambda U ^\T-\rho^{-1} U \Lambda \Lambda_1 U ^\T+I_p-U \Lambda_1 U ^\T\\
=&I_p +\rho^{-1} U (\Lambda-  \Lambda \Lambda_1-\rho \Lambda_1)U ^\T=I_p.
\end{align*}
For the second formula \eqref{f2}, we evaluate the four parts on the right hand side respectively. Firstly we have,  
\begin{align}
&(2^{-1}S \otimes I_p+2^{-1}I_p \otimes S+\rho I_{p^2})\rho^{-1} I_{p^2} \nonumber\\
=& I_{p^2}+(2 \rho)^{-1}(U \Lambda U^\T) \otimes I_p+(2 \rho)^{-1} I_p \otimes (U \Lambda U^\T). \label{a1}
\end{align}
Secondly, 
\begin{align}
&(2^{-1}S \otimes I_p+2^{-1}I_p \otimes S+\rho I_{p^2})\{-\rho^{-1} (U \Lambda_2 U^\T) \otimes I_p\}\nonumber \\
=&-(2 \rho)^{-1}(U \Lambda \Lambda_2 U^\T) \otimes I_p-(2 \rho)^{-1}(U \Lambda_2 U^\T) \otimes (U \Lambda U^\T)-(U \Lambda_2 U^\T) \otimes I_p \nonumber \\
=&-(2 \rho)^{-1} \{U (\Lambda \Lambda_2+2 \rho \Lambda_2) U^\T  \} \otimes I_p-(2 \rho)^{-1}(U \otimes U)(\Lambda_2 \otimes \Lambda)(U^\T \otimes U^\T ) \nonumber \\
=&-(2 \rho)^{-1} (U \Lambda U^\T)\otimes I_p-(2 \rho)^{-1}(U \otimes U)(\Lambda_2 \otimes \Lambda)(U^\T \otimes U^\T ). \label{a2}
\end{align}
Thirdly, similarly to \eqref{a2}, we have
\begin{align}
&(2^{-1}S \otimes I_p+2^{-1}I_p \otimes S+\rho I_{p^2})\{-\rho^{-1} I_p \otimes  (U \Lambda_2 U^\T) \}\nonumber \\
=&-(2 \rho)^{-1} I_p \otimes  (U \Lambda U^\T)-(2 \rho)^{-1}(U \otimes U)(\Lambda \otimes \Lambda_2)(U^\T \otimes U^\T ), \label{a3}
\end{align}
and lastly, we have
\begin{align}
&(2^{-1}S \otimes I_p+2^{-1}I_p \otimes S+\rho I_{p^2})\{\rho^{-1} (U \otimes U) \diag\{\vec(\Lambda_3)\}(U^\T \otimes U^\T )\}\nonumber \\
=&(2\rho)^{-1} (U \otimes U) (\Lambda \otimes I_m) \diag\{\vec(\Lambda_3)\}(U^\T \otimes U^\T )\nonumber \\
&+(2\rho)^{-1} (U \otimes U) (I_m \otimes \Lambda) \diag\{\vec(\Lambda_3)\}(U^\T \otimes U^\T )\nonumber \\
&+(U \otimes U) \diag\{\vec(\Lambda_3)\}(U^\T \otimes U^\T ) \nonumber \\
=& (2\rho)^{-1} (U \otimes U) \diag\{\vec(\Lambda_3 \Lambda+\Lambda \Lambda_3+2\rho \Lambda_{3})\}(U^\T \otimes U^\T ). \label{a4}
\end{align}
Combing \eqref{a1}, \eqref{a2}, \eqref{a3} and \eqref{a4}, it suffices to show 
\begin{align*}
\diag\{\vec(\Lambda_3 \Lambda+\Lambda \Lambda_3+2\rho \Lambda_{3})\}=\Lambda_2 \otimes \Lambda+\Lambda \otimes \Lambda_2,
\end{align*}
which is true since 
\begin{align*}
\Lambda_3 \Lambda+\Lambda \Lambda_3+2\rho \Lambda_{3}=& \left\{\frac{\tau_i \tau_j(\tau_i+\tau_j+4\rho)}{(\tau_i+2\rho)(\tau_j+2\rho)}\right\}_{m \times m}=\left\{\frac{\tau_i \tau_j}{\tau_i+2\rho}+\frac{\tau_i \tau_j}{\tau_j+2\rho}\right\}_{m \times m}.
\end{align*}
The proof is completed.  \hfill$\fbox{}$
\subsection{Proofs of Theorem \ref{prop1}}
Conclusion (i) is a direct result of Proposition \ref{prop0}, and next we provide proofs for conclusion (ii). Note that
\begin{align*}
\vec(2^{-1}S \Omega +2^{-1}\Omega S+\rho \Omega)=(2 ^{-1}  I_p \otimes S +2^{-1} S \otimes I_p+\rho I_{p^2}) \vec(\Omega).
\end{align*}
Therefore, the solution is given by
\begin{align*}
\vec(\Omega)=&(2^{-1}S \otimes I_p+2^{-1} I_p \otimes S+\rho I_{p^2})^{-1} \vec(C).
\end{align*}
By Proposition \ref{prop0}, 

\begin{align*}
\vec(\Omega)=&\{\rho^{-1} I_{p^2}-\rho^{-1} (U \Lambda_2 U^\T) \otimes I_p-\rho^{-1} I_p \otimes (U \Lambda_2 U^\T)\\
&+\rho^{-1} (U \otimes U) \diag\{\vec(\Lambda_3)\}(U^\T \otimes U^\T ) \}  \vec(C)\\
=& \rho^{-1} \vec(C)-\rho^{-1}  \vec(C U \Lambda_2 U^\T)-\rho^{-1}  \vec(U \Lambda_2 U^\T C)\\
&+\rho^{-1}(U \otimes U) \diag\{\vec(\Lambda_3)\} \vec(U^\T C U)\\
=& \rho^{-1} \vec(C)-\rho^{-1}  \vec(C U \Lambda_2 U^\T)-\rho^{-1}  \vec(U \Lambda_2 U^\T C)\\
&+\rho^{-1}(U \otimes U) \vec\{\Lambda_3 \circ (U ^\T C U)  \} \\
=& \rho^{-1} \vec(C)-\rho^{-1}  \vec(C U \Lambda_2 U^\T)-\rho^{-1}  \vec(U \Lambda_2 U^\T C)\\
&+\rho^{-1} \vec\{ U  [\Lambda_3 \circ (U ^\T C U)]U^\T  \} \\
\end{align*}
which yields 
\begin{align*}
	\Omega=\rho^{-1}C-\rho^{-1}C U \Lambda_2 U  ^\T- \rho^{-1}U \Lambda_2 U ^\T C+\rho^{-1} U \{\Lambda_3 \circ (U ^\T C U)\} U ^\T.
\end{align*}
The proof is completed.  \hfill$\fbox{}$

\section*{References}

\bibliographystyle{apalike}
\bibliography{ref}
\end{document}